\documentclass[12pt,a4paper]{article}

\usepackage{authblk}

\usepackage{graphicx}
\usepackage{lmodern}
\usepackage[T1]{fontenc}
\usepackage{amsmath,amssymb}
\usepackage{cite}

\usepackage[left=2cm,right=2cm,top=2.5cm,bottom=2.5cm]{geometry}

\usepackage[font={footnotesize}]{caption} %%% without setspace
\usepackage[rubberchapters]{titlesec}
\titleformat{\subsection}[hang]
{\bfseries}
{\normalsize\thesubsection}
{10pt}
{\normalsize}
[]
\titlespacing*{\subsection}{0cm}{0.6cm}{0.2cm}[0.5cm]

\titleformat{\section}[hang]
{\bfseries}
{\large\thesection}
{10pt}
{\large}
\titlespacing*{\section}{0.0cm}{1.0cm}{0.3cm}[0.5cm]

\usepackage[colorlinks=true,urlcolor=blue,linktoc=page,linkcolor=blue,citecolor=blue]{hyperref}

\newcommand{\emailcom}[1]{\href{mailto:#1}{\texttt{#1}}}

\numberwithin{equation}{section}
\newcommand{\authcom}{\normalsize\bfseries\sffamily}
\author{\authcom M.~Albaladejo\thanks{\emailcom{Miguel.Albaladejo@ific.uv.es}}}
\author{C.~Hidalgo-Duque\thanks{\emailcom{Carlos.Hidalgo@ific.uv.es}}}
\author{J.~Nieves}
\author{E.~Oset \thanks{\emailcom{Eulogio.Oset@ific.uv.es}}}
\affil[]{{\normalsize\it Instituto de F\'isica Corpuscular (centro mixto CSIC-UV), Institutos de Investigaci\'on de Paterna, Aptdo.~22085, 46071, Valencia, Spain}}

\title{\bfseries\sffamily Hidden charm molecules in finite volume}
\date{\normalsize\today}
\begin{document}
\maketitle

\begin{abstract}
In the present paper we address the interaction of pairs of charmed mesons with hidden charm in a finite box. We use the interaction from a recent model based on heavy quark spin symmetry that predicts molecules of hidden charm in the infinite volume. The energy levels in the box are generated within this model, and from them some synthetic data are generated. These data are then employed to study the inverse problem of getting the energies of the bound states and phase shifts for $D \bar D$ or $D^* {\bar D}^*$. Different strategies are investigated using the lowest two levels for different values of the box size, carrying a study of the errors produced. Starting from the upper level, fits to the synthetic data are carried out to determine the scattering length and effective range plus the binding energy of the ground state. A similar strategy using the effective range formula is considered with a simultaneous fit to the two levels, one above and the other one below threshold. This method turns out to be more efficient than the other one. Finally, a method based on the fit to the data by means of a potential and a loop function conveniently regularized, turns out to be very efficient and allows to produce accurate results in the infinite volume starting from levels of the box with errors far larger than the uncertainties obtained in the final results. A regularization method based on Gaussian wave functions turns out to be rather efficient in the analysis and as a byproduct a practical and fast method to calculate the L\"uscher function with high precision is presented.
\end{abstract}

\section{Introduction}
The determination of the hadron spectrum from lattice QCD (LQCD) calculations
is attracting many efforts and one can get an overview on the
different methods used and results in the recent review
\cite{reviewhadron}. One of the tools becoming gradually more used is
the analysis of lattice levels in terms of the L\"uscher method
\cite{Luscher:1986pf,Luscher:1990ux}. This method converts binding energies
of a hadron-hadron system in the finite box into phase shifts of
hadron-hadron interaction from levels above threshold, or binding
energies from levels below
threshold~\cite{Luscher:1985dn,Beane:2003da, Beane:2011iw}. From the
phase shifts one can get resonance properties, and there are several
works that have recently applied these techniques to study the $\rho$
resonance~\cite{Feng:2011ah,Aoki:2007rd,
  Gockeler:2008kc,Aoki:2010hn,Feng:2010es,Frison:2010ws,Lang:2011mn,Prelovsek:2011im, dudek}.
There exist other resonances far more difficult to get with this
approach like the $a_1(1260)$, which was also attempted in
\cite{Prelovsek:2011im} (see also its determination using finite
volume calculations with effective field theory in
\cite{Roca:2012rx}). Scalar mesons have also been searched for with
this method \cite{Prelovsek:2010kg,Alexandrou:2012rm, fu1, fu2} and gradually
some calculations are being performed for systems in the charm sector
\cite{Mohler:2011ke,Kalinowski:2012re,Mohler:2012na,Ozaki:2012jv,
  Kawanai:2010ru,Yang:2012mya, guo}. From another field theoretical perspective,
finite volume calculations have also been devoted to this sector in
\cite{koren,xie, guo}. In \cite{koren} the $\bar KD$, $\eta D_s$
interaction is studied in finite volume with the aim of learning about
the nature of the $D_{s0}^{*} (2317)$ resonance from lattice
data.\footnote{The first hint, though indirect, of the nature of the
  $D_{s0}^{*} (2317)$ as a mostly $\bar K D$ bound state from lattice
  data was presented in \cite{Flynn:2007ki}. There, lattice
  calculations of the scalar form factors in semileptonic
  pseudoscalar-to-pseudoscalar decays were used to extract
  information about the corresponding elastic $S$-wave scattering
  channels.} The  theoretical model used in \cite{koren} is
taken from \cite{Gamermann:2006nm}, where the $D_{s0}^{*}(2317)$
resonance appears dynamically generated from the interaction of $\bar
KD$, $\eta D_s$ and other less relevant channels. In this latter work, a scalar hidden charm state coming from the $D \bar
D$ interaction with other coupled channels was also found, which qualifies basically
as a $D \bar D$ quasi-bound state (decaying into pairs of lighter
pseudoscalars). Although not reported experimentally, support for this
state has been found in \cite{Gamermann:2007mu} from the analysis of
the data of the $e^+ e^- \to J / \psi D \bar D$ reaction of
\cite{Abe:2007sya}. From the effective field theory point of view, this
state has also been reported in \cite{Nieves:2012tt,HidalgoDuque:2012pq}, using
light SU(3)-flavour and heavy quark spin symmetries to describe charmed
meson-antimeson interactions.

The purpose of the present paper is to study the interaction of $D
\bar D$ and $D^* {\bar D}^*$ using a field theoretical approach in
finite volume in order to evaluate energy levels in the finite box
which might be compared with future LQCD calculations. The
paper also presents a strategy to better analyze future lattice
results in order to get the best information possible about bound
states and phase shifts in the infinite volume case from these lattice
data.  For this purpose we shall use the model of
\cite{HidalgoDuque:2012pq}, although most of the results and the basic
conclusions are independent of which model is used.
  
  As to the method to obtain the finite volume levels and the inverse
  problem of obtaining the results in the real world, phase shifts and
  binding energies, we shall follow the method of \cite{Doring:2011vk} where a
  reformulation of L\"uscher approach is done based on the on shell
  factorization of the scattering matrix that one uses in the chiral
  unitary approach \cite{aoorev,Nieves:1998hp,Nieves:1999bx,Oller:2000fj, hyodo, carmen}. This method
  is conceptually and technically very easy and introduces
  improvements for the case of relativistic particles (although we
  shall not make use of the relativistic version in the present
  paper). Some works using this formalism can be found in Refs.~\cite{koren, Roca:2012rx, xie, Doring:2011ip, Doring:2011nd, Chen:2012rp, MartinezTorres:2012yi, Albaladejo:2012jr}.

\section{Formalism: infinite volume}
\label{sec:infinite_volume}
In this section, we briefly review the formalism of
Refs.~\cite{Nieves:2012tt,HidalgoDuque:2012pq}, where an effective field theory
incorporating SU(3)-light flavour symmetry and heavy quark spin symmetry (HQSS) is
formulated, to study charmed meson-antimeson (generically denoted here
$H\bar{H}'$, with $H,H'=D,D^\ast, D_s$, $D^\ast_s$) bound states. The
lowest order (LO) contribution of the interaction is given by contact
terms, and the symmetries reduce the number of independent low energy constants (LECs)
of the approach to only four. Other effects, like one-pion exchange or
coupled channel dynamics, are shown to be sub-leading corrections to
this order. Still, coupled channels will be considered explicitly when
the mass difference between the thresholds is not negligible compared
with the binding energy of the molecules considered. To fix the four
constants of the approach, one assumes the molecular nature of some
XYZ states, namely, $X(3872)$, $X(3915)$ and $Y(4140)$. The fourth
input of the model is the isospin violating branching ratio of the
decays $X(3872)\to J/\Psi \omega$ and $X(3872)\to J/\Psi \rho$ (for a
different approach to this issue and the $X(3872)\to J/\Psi \gamma$
decay see \cite{fca}). For further details on the formalism we refer
to
Refs.~\cite{HidalgoDuque:2012pq,Valderrama:2012jv,Nieves:2012tt}. We
will adapt here the formalism to a more adequate (for the problem in
hands) $T$-matrix language.

Since we are dealing with heavy mesons, we use a non-relativistic
formalism. In our normalization, the $S$-matrix\footnote{We will
  always consider $S-$wave meson-antimeson interactions, and thus the
  spin of the molecule will always coincide with the total spin of the
meson-antimeson pair. The partial waves $^{2S+1}L_J$ are then
$^{2S+1}S_{J=S}$. For simplicity in what follows,  we will drop all
references to the $L$, $S$ and $J$ quantum numbers, both in the $S$
and $T$ matrices.} for an elastic
$H\bar{H}'$ scattering process reads
\begin{equation}
\label{eq:smatrix}
S(E) \equiv e^{2i\delta(E)} = 1-i\frac{\mu k}{\pi}T(E)~,
\end{equation}
where the modulus of the momentum $k=\lvert \vec{k} \rvert$ is given by $k^2 =
2\mu(E-m_1-m_2)$, and $\mu$ is the reduced mass of the system of
two particles with masses $m_1$ and $m_2$. In Eq.~\eqref{eq:smatrix}, $\delta$ is the phase shift, and we can write:
\begin{align}
T & = -\frac{2\pi}{\mu k} \sin\delta e^{i\delta}~,\label{eq:tmat_phase}\\
T^{-1} & = - \frac{\mu k}{2\pi} \cot\delta + i \frac{\mu k}{2\pi}~.\label{eq:tmat_kcotd}
\end{align}
The expression for the $T$-matrix is given by:
\begin{equation}
\label{eq:tmat_ND}
T^{-1}(E) = V^{-1}(E) - G(E)~,
\end{equation}
%  %The loop function (potential) $G$ ($V$) provides the right (left) hand cut of the $T-$matrix. 
with $V$ the potential (two-particle irreducible amplitude) and $G$
a one-loop two-point function. This equation stems from a  once-subtracted dispersive
representation of $T^{-1}(E)$ (see for instance Sec. 6 of
Ref.~\cite{Nieves:1999bx}), or equivalently, from the $N/D$
method \cite{Chew:1960iv} equations, when the left-hand cut is
neglected or included perturbatively \cite{Oller:2000fj, Oller:1998zr,  
Albaladejo:2008qa, Albaladejo:2012te}. The loop function $G$ provides the right-hand cut and the contribution of the left-hand cut should be included in the potential $V$. As mentioned above, we will follow here the approach of
Refs.~\cite{Nieves:2012tt,HidalgoDuque:2012pq}, and we will
approximate $V$ by its LO contribution in the $1/m_Q$ expansion (with
$m_Q$ the mass of the heavy quark). Thus, we are completely
neglecting the left-hand cut.

The loop function $G$ needs to be regularized in some way. Typical
approaches are once-subtracted dispersion relations and sharp
cutoffs. Here, instead, we are following the approach of
Refs.~\cite{Nieves:2012tt,HidalgoDuque:2012pq}, in which the loop
function is regularized with a Gaussian regulator. For an arbitrary
energy $E$, we find 
\begin{align}
G(E) & = 
\int \frac{\text{d}^3 \vec{q}}{(2\pi)^3} \frac{e^{-2(\vec{q}^{\,2}-k^2)/\Lambda^2}}{E-m_1-m_2 - \vec{q}^{\,\,2}/2\mu + i0^+} \nonumber\\
& = -\frac{\mu\Lambda}{(2\pi)^{3/2}}e^{2k^2/\Lambda^2} + \frac{\mu k}{\pi^{3/2}} \phi\left(\sqrt{2}k/\Lambda\right)-i \frac{\mu k}{2\pi}~,\label{eq:gmat_gr}
\end{align}
with $\phi(x)$ given by:
\begin{equation}
\phi(x) = \int_{0}^{x} e^{y^2} \text{d}y~.
\end{equation}
Note that, the wave number
$k$ is a multivalued function of $E$, with a branch point at
threshold ($E=m_1+m_2$). The principal argument of $(E-m_1-m_2)$
should be taken in the range $[0,2\pi[$.  The function $k
    \phi(\sqrt2 k/\Lambda)$ does not present any discontinuity
    for real $E$ above threshold, and $G(E)$ becomes a multivalued
    function because of the $ i k $ term. Indeed, $G(E)$ has two
    Riemann sheets. In the first one, $0\leqslant {\rm
      Arg}(E-m_1-m_2)< 2\pi$, we find a discontinuity $G_I(E+ i\epsilon)-G_I(E-i \epsilon) = 2i\,{\rm Im} G_I(E+i\epsilon)$
    for $E> (m_1+m_2)$. It guaranties that the $T-$matrix fulfills the
    optical theorem.  For real values of $E$ and below threshold,   we have 
$k = i\,\sqrt{-2\mu (E-m_1-m_2)}$. Poles below
    threshold in the first sheet correspond to bound states. In the
    second Riemann sheet, $2\pi\leqslant {\rm Arg}(E-m_1-m_2)<
  4\pi$, we trivially find $G_{II}(E- i\epsilon) = G_I(E+i\epsilon)$, for real energies and above threshold. 
  
The Gaussian form factor enters Eq.~\eqref{eq:gmat_gr} in a way that is unity for on-shell momenta, and hence the optical theorem $\text{Im} T^{-1} = \mu k/(2\pi)$ is automatically fulfilled. A Gaussian regulator is also used in Ref.~\cite{tony} to study the $\Delta$ resonance in finite volume.

The cutoff $\Lambda$ is a parameter of the approach, and, hence, the
theory depends on it. This dependence, however, is partially
reabsorbed in the counter-terms of the theory, as long as one chooses a
reasonable value for it, not beyond the high-energy scale of the
effective field theory \cite{Lepage:1989hf, Lepage:1997cs, Epelbaum:2006pt,
  Epelbaum:2009sd}.

In the approach of Refs.~\cite{Nieves:2012tt,HidalgoDuque:2012pq}, 
the potential $V$ is taken as
\begin{equation}\label{eq:potential_gr}
V(E) = e^{-2k^2/\Lambda^2} C(\Lambda)~,
\end{equation}
where $C$ is the proper combination of the four different
counter-terms for each considered channel $H\bar{H}'$. Explicit
expressions can be found in
Appendix~\ref{app:potentials_constants}. The dependence of the
counter-term on the ultraviolet (UV) cutoff $\Lambda$ should cancel that
of the loop function $G$, such that $G(E_B) V(E_B)$ becomes
independent of $\Lambda$, when $E_B$ is the energy of the bound state
used to determine the counter-term.\footnote{The mass of a bound state
  is thus given by $T^{-1}(E_B)=0$ for $E_B<m_1+m_2$, which is the
  equivalent of Eq.~(10) in Ref.~\cite{HidalgoDuque:2012pq}.} For
other energies, there will exist a remaining, unwanted/unphysical,
dependence of the $T$ matrix on the cutoff. This is due to the
truncation of the perturbative expansion (see discussion in
Ref.~\cite{Nieves:2012tt}).  Up to this point, we have discussed only
the case of uncoupled channels, but the generalization to coupled
channels is straightforward.\footnote{One just has to rewrite the
  $T$-matrix as $T=(\mathbb{I}-V G)^{-1} V$, where $V$ and $G$ are now
  matrices in the coupled channels space.}

Finally, above threshold the effective range expansion reads:
\begin{equation}\label{eq:eff_range}
k \cot \delta = -\frac{1}{a} + \frac{1}{2} r k^2 + \cdots~,
\end{equation}
where $a$ and $r$ are, respectively, the scattering length and the
effective range. From Eqs.~\eqref{eq:tmat_kcotd} and
\eqref{eq:tmat_ND} we can calculate the theoretical predictions for
these effective range parameters, obtaining:

\begin{align}
a_\text{th} & = \hphantom{-}\frac{\mu}{2\pi} \left( \frac{1}{C} +
 \frac{\mu \Lambda}{(2\pi)^{3/2}} \right)^{-1}~,\label{eq:sl_th}\\
r_\text{th} & =-\frac{8\pi}{\mu \Lambda^2} \left( \frac{1}{C} - 
\frac{\mu \Lambda}{(2\pi)^{3/2}}\right).\label{eq:er_th}
\end{align}

%%%%%%%%%%%%%%%%%%%%%%%%%%%%%%%%%%%%%%%%%%%%%%%%%%
\section{Formalism: finite volume}
\label{sec:FV}
In this section, we follow the steps of Ref.~\cite{Doring:2011vk} to
write the amplitude in a finite box of size $L$ with periodic boundary
conditions, denoted by $\tilde{T}$. Since the potential does not
depend on $L$, one only has to replace the loop function $G$ with its
finite volume version, $\tilde{G}$, in which the integral over
momentum $\vec{q}$ is replaced by a discrete sum over the allowed
momenta,
\begin{align}
\tilde{T}^{-1}(E) & = V^{-1}(E) - \tilde{G}(E)~,\label{eq:tmat_FV}\\
\tilde{G}(E) & = \frac{1}{L^3}\sum_{\vec{q}} \frac{e^{-2(\vec{q}^{\,2}-k^2)/\Lambda^2}}{E-m_1-m_2 - \vec{q}^{\,2}/2\mu}~,\label{eq:gfun_FV}
\end{align}
where the (quantized) momentum is given by:
\begin{equation}
\vec{q} = \frac{2\pi}{L}\vec{n}~,\quad \vec{n} \in \mathbb{Z}^3~.
\end{equation}
Now, the energy levels in the box are given by the poles of the
$\tilde{T}$-matrix, $V^{-1} = \tilde{G}$. For the energies of these levels in the
box, the amplitude in the infinite volume is recovered as:
\begin{equation}
\label{eq:euluscher}
T^{-1}(E) = V^{-1}(E)-G(E) = \tilde{G}(E) - G(E) = \delta G(E)~.
\end{equation}
Since the $G$ function is regularized (either in the box or in the
infinite volume) with a Gaussian regulator, the difference above
depends explicitly on the cutoff $\Lambda$. This remaining non-physical 
dependence on $\Lambda$ quickly disappears as the volume increases. Indeed, we
find that it is exponentially suppressed and that it dies off as
$\exp{(-L^2\Lambda^2/8)}$ (see Appendix \ref{app:limit}). Thus, it is clear that in this context, we can
 end up the renormalization program just by sending the UV
 cutoff to infinity. This will allow  to obtain the physical $T-$matrix, independent of any renormalization scale, for the energy
 levels found in the lattice Monte Carlo simulation (finite box).

For the practical calculations that we will show in what follows, the $\Lambda$ dependence
is already negligible when $\Lambda \gtrsim 1\ \text{GeV}$ even for
the smallest volumes considered in this work (the limit $\Lambda \to
\infty$ is effectively achieved for such values). In this limit,
Eq.~\eqref{eq:euluscher} becomes the L\"uscher equation
\cite{Luscher:1986pf, Luscher:1990ux}, as we discuss in certain detail in
Appendix \ref{app:limit}.  The results of the appendix also show that the inclusion of a Gaussian regulator is a quite efficient technique, from the
computational point of view, to evaluate the L\"uscher
function ${\cal Z}_{00} (1,\hat k^2)$ used in \cite{Luscher:1990ux}. 
Finally, from Eqs.~\eqref{eq:tmat_kcotd} and \eqref{eq:euluscher}, we
can write:
\begin{equation}\label{eq:kcotd_luscher}
k\cot\delta = -\frac{2\pi}{\mu} \lim_{\Lambda \to \infty} \text{Re}\left(\tilde{G}(E)-G(E) \right)~.
\end{equation}
%
%%%%%%%%%%%%%%%%%%%%%%%%%%%%%%%%%%%%%%%%%%%%%%%%%%%
\section{Results}
We present in this section the results obtained with the formalism
outlined in the previous section. We first discuss the results
obtained by putting the model of
Refs.~\cite{Nieves:2012tt,HidalgoDuque:2012pq} directly in 
the box. That is, we study
the volume dependence of the molecules found in
Ref.~\cite{HidalgoDuque:2012pq}, predicting thus the existence of
sub-threshold levels (asymptotically different of threshold) for the
different channels, which have a clear correspondence with the
hidden charm molecules reported in  \cite{HidalgoDuque:2012pq}. This is done in
Subsec.~\ref{subsec:results_model_box}.

Our purpose in Subsecs.~\ref{subsec:phase}, \ref{subsec:potential} and \ref{subsec:neweffectiverange}
is to simulate a realistic situation in a LQCD study, where one would
obtain different energy levels (one or two) for different sizes, $L$,
of the box. To do so, we generate ``synthetic data'' from the exact
levels that we obtain from the model of
Refs.~\cite{Nieves:2012tt,HidalgoDuque:2012pq}. We take five different
values of $L_i$, in the range $L m_\pi = 1.5$ to $3.5$. From the
calculated levels, we obtain randomly shifted levels (in a range of
$5\ \text{MeV}$), and assign an error of $10\ \text{MeV}$ to each of
these points, except the last one for which we have assumed an error
of only $8\ \text{MeV}$ to prevent it from crossing the threshold. Next we use a Monte
Carlo simulation, to estimate the errors on the determination of observables
(the phase shifts, for instance) when the energy levels are obtained 
with a certain statistical error. Specifically, we study in these subsections the $I=0$ $J^{PC}=0^{++}$ $D\bar{D}$ channel.

In Subsec.~\ref{subsec:phase}, the L\"uscher formalism to study the
phase shifts calculated from Eq.~\eqref{eq:kcotd_luscher} is applied to the
synthetic levels above threshold that we find for the different studied channels. From these phase shifts, we calculate the effective range
expansion parameters, and use them to determine the masses of the bound
states. In Subsec.~\ref{subsec:potential} we adopt another strategy
to extract information from the generated levels. Namely, we consider
a potential which parameters are then fitted to reproduce the
synthetic levels (above and below threshold, simultaneously). With this potential, we
can make predictions in the infinite volume case, and thus we end up with another
determination of the masses of the predicted bound states. We shall see that this method allows one to obtain better results (better central value and smaller errors) for the mass of the bound state than the previous one. We then analyze in detail which are the differences of both approaches. In Subsec.~\ref{subsec:neweffectiverange} another method is proposed, in which the effective range approximation is retained for the inverse of the $T$-matrix amplitude, but fitting directly the energy levels instead of the phase shifts, and studying simultaneously the levels above and below threshold. In this case, then, we notice that the precision achieved for the mass of the bound state is similar to that obtained with the potential analysis.

In Subsec.~\ref{subsec:bound_state_fit}, we analyze in a
more quantitative way, the qualitative arguments given in Subsec.~\ref{subsec:results_model_box}, where the behavior of the sub-threshold levels is discussed. We offer a method to discriminate
between those levels that produce bound states in the limit $L\to \infty$
and those that do not, and hence tend to threshold in the
infinite volume limit. This method allows the extraction of the mass and the
coupling of the bound state in the infinite volume limit.

All these methods are applied in Subsec.~\ref{subsec:2plusplus} to the bound state present in the $I=0$ $J^{PC}=2^{++}$ $D^\ast \bar{D}^\ast$ channel. The difference with respect to the case used as an example in the previous subsections is that the state is now weakly bound (the binding energy is only around $2$--$3\ \text{MeV}$), so that we can compare how the methods exposed above work for this case.

\subsection{The model of Ref.~\cite{HidalgoDuque:2012pq} in a finite box}
\label{subsec:results_model_box}
\begin{figure}[ht]
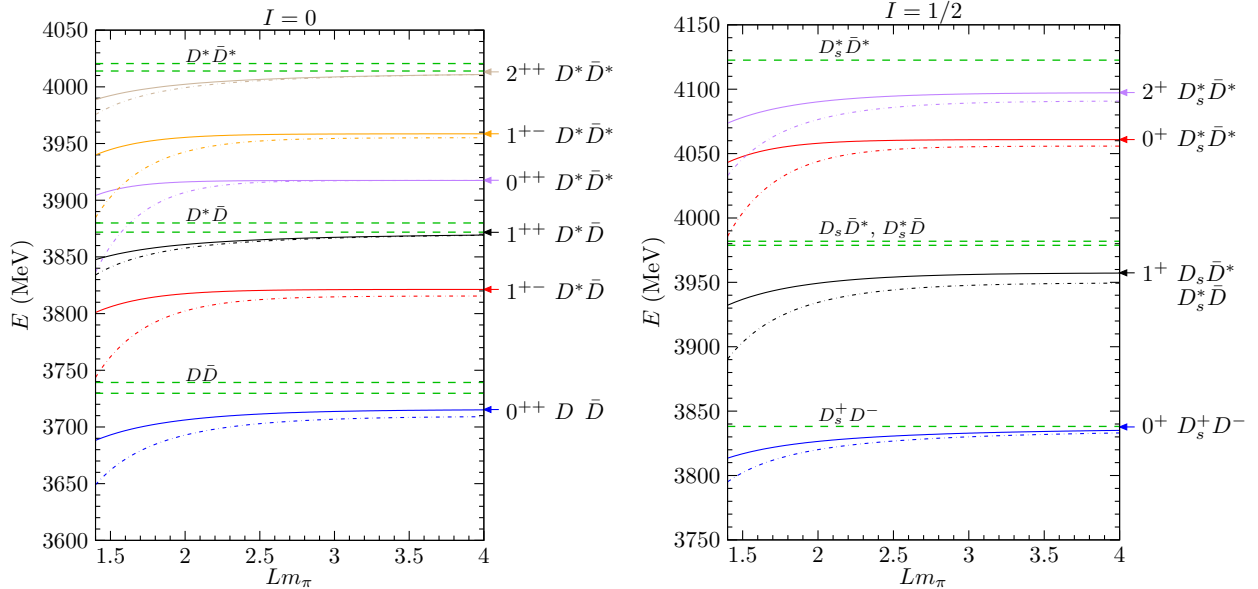
\centering
\includegraphics[width=0.47\textwidth,keepaspectratio]{BoundStatesFigures_0.mps}\hspace{5pt}
\includegraphics[width=0.47\textwidth,keepaspectratio]{BoundStatesFigures_1.mps}
\caption{Volume dependence of the $I=0$ (left) and $I=1/2$ (right)
  molecules predicted in Ref.~\cite{HidalgoDuque:2012pq}. The
  horizontal dashed lines show the different thresholds  
  involved (when the charge is not explicitly given, we are
  displaying the thresholds associated to the different charge channels). The solid lines correspond to the levels found in the box
  for $\Lambda=1\ \text{GeV}$, whereas the dot-dashed ones stand for those obtained with $\Lambda=0.5\ \text{GeV}$. Over the right axis we mark with
  arrows the masses of the bound states as predicted in the infinite
  volume case and $\Lambda=1\ \text{GeV}$. 
  The $J^{PC}$ quantum numbers of the different channels are
  indicated beside the arrows.\label{fig:BS_from_model_1}}
\end{figure}
\begin{figure}[ht]
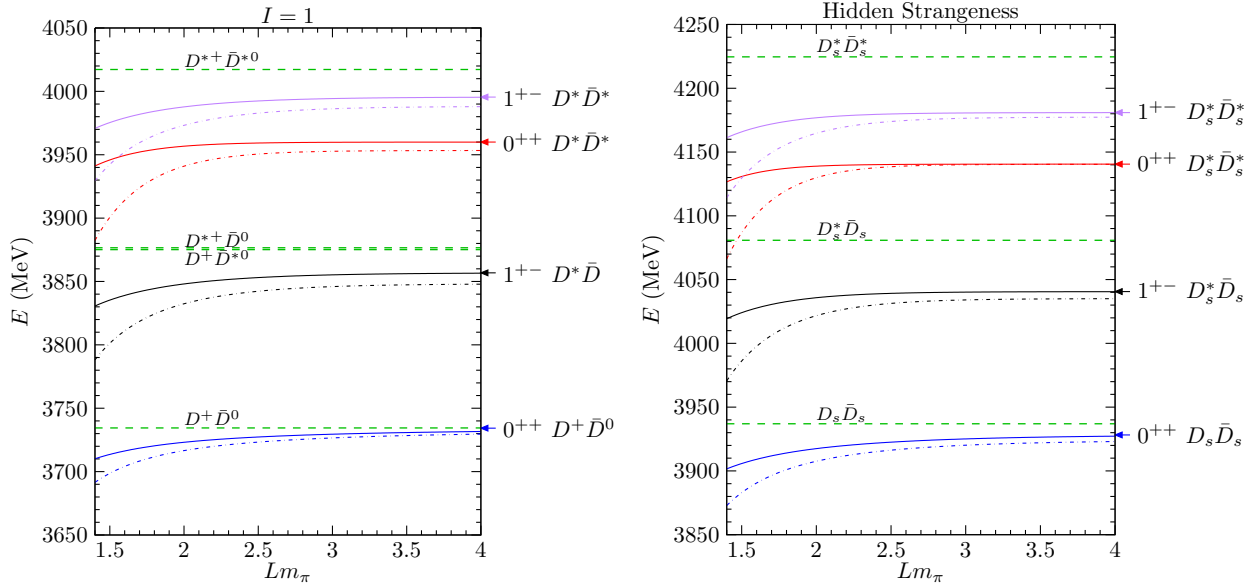
\centering
\includegraphics[width=0.47\textwidth,keepaspectratio]{BoundStatesFigures_2.mps}\hspace{5pt}
\includegraphics[width=0.47\textwidth,keepaspectratio]{BoundStatesFigures_3.mps}
\caption{The same as in Fig.~\ref{fig:BS_from_model_1} for the $I=1$
  (left) and hidden strangeness (right) molecules predicted in
  Ref.~\cite{HidalgoDuque:2012pq}. \label{fig:BS_from_model_2}}
\end{figure}

In Figs.~\ref{fig:BS_from_model_1} and \ref{fig:BS_from_model_2} we
present the dependence of the energy levels on the size of the
finite box, as calculated from Eqs.~\eqref{eq:tmat_FV} and \eqref{eq:gfun_FV}, for the different channels studied in
Ref.~\cite{HidalgoDuque:2012pq}. We have fixed the potential in the
different channels by means of the central values given in this
reference for the various counter-terms, and collected here in Appendix \ref{app:potentials_constants}, Eqs.~\eqref{eq:c0a}--\eqref{eq:c1b}. When needed, we have also
implemented in the finite box a coupled channel formalism.  The solid
lines correspond to the case $\Lambda = 1\ \text{GeV}$, whereas the
dot-dashed lines to $\Lambda= 0.5\ \text{GeV}$. For comparison, we
also show, with the horizontal dashed lines, the involved threshold
energies. We just show those energy levels that can be identified with
bound states ($k^2 < 0$) in the infinite volume case. That is, their
asymptotic $L\to\infty$ value approaches the bound energies given
in \cite{HidalgoDuque:2012pq}, and thus they are different from
threshold. Of course, one has this latter piece of information from
the calculations of the model in an infinite volume, but this would
not be the case in a lattice simulation. Let us focus, for simplicity,
in the $I=0$ case, shown in the left panel of
Fig.~\ref{fig:BS_from_model_1}. The large $L$ asymptotic behavior
can be well appreciated in some cases like the $0^{++}$ $D\bar{D}$ or
$1^{+-}$ $D^\ast \bar{D}$ molecules. However in other cases, it might
be difficult to discriminate between a real bound state and a
threshold level, even for quite large values of the box size
$L$. Clear examples are the $1^{++}$ $D^\ast \bar{D}$ or the $2^{++}$
$D^\ast \bar{D}^\ast$ molecules (similar examples can be found in the
different isospin-strangeness channels), which in the infinite volume
case are loosely bound.\footnote{The first one corresponds to the
  $X(3872)$ resonance that has been observed close to the
  $D^0 \bar{D}^{0\ast}$ threshold~\cite{Choi:2003ue} (see also a recent determination and discussion of other experiments in Ref.~\cite{kamal}) and it has been a hot
  topic for both the experimental and theory communities since its
  discovery.  The $2^{++}$ state is a HQSS
  partner of the $X(3872)$ molecule which dynamics, at LO in the heavy
  quark expansion, is being determined by precisely the same
  combination of counter-terms that appear in the $X(3872)$
  channel. Given the discovery of the $X(3872)$ resonance, the
  existence of the $2^{++}$ state, either as a bound state or a
  resonance, is therefore a quite robust consequence of HQSS~\cite{Nieves:2012tt,HidalgoDuque:2012pq}.}  Thus, we
see a well known result from Quantum Mechanics; the smaller the binding energies, the larger
become the $L$ values needed to reach the
asymptotic behavior. From this study, we conclude that 
 in a lattice simulation when dealing with states that are at
 least bound by some tens of MeV, one might 
safely discriminate them by using box sizes of  
the order of $Lm_\pi \simeq 3$. However, the lattice study of 
other states less bound (binding  energies below 4 or 5 MeV, as in 
the case of the $X(3872)$ or the $2^{++}$ resonances) might require
significantly larger volumes.  To achieve more accurate predictions for
 the former and solve the problem for the
latter ones, we follow different approaches in the following
subsections.

Finally, we note that some of the levels in Figs.~\ref{fig:BS_from_model_1} and \ref{fig:BS_from_model_2} are not
realistic, in the sense that they would mix with other levels
generated by channels with the same quantum numbers, but lower
thresholds. That is the case, for example in $I=0$, of the $0^{++}$
$D^\ast \bar{D}^\ast$ at $E\simeq 3920\ \text{MeV}$, that would mix
with some higher levels of the $D \bar{D}$ channel. Indeed, it
is to be expected that these bound states would acquire some width due
to the coupled channel dynamics. Still, it is possible that these
states could appear as more or less stable energy levels.\footnote{In Ref.~\cite{raquel} the $D^\ast \bar{D}^\ast$, $D^\ast_s \bar{D}^\ast_s$ states are studied with the interaction taken from the extrapolation of the local hidden gauge approach to the charm sector, which also respects HQSS. The coupling to $D\bar{D}$ and $D_s \bar{D}_s$ is allowed and generates a width of about $50\ \text{MeV}$ for the most bound state, the one with $I=0$ and $J^{PC}=2^{++}$.}

%Also,
%let us note that in other models that respect heavy quark spin
%symmetry, based on dynamics of the local hidden gauge approach
%\cite{hidden1, hidden2, hidden4}, the transition from $D$ to $D^*$
%is based on $\pi$ exchange that is shown to be subleading in the heavy
%mass counting \cite{wu, wu2, wub, xiaojuan}. The states with $D$ and
%$D^\star$ then barely mix.

\subsection{Inverse analysis: phase shifts}
\label{subsec:phase}
\begin{figure}[ht]\centering
\includegraphics[width=0.6\textwidth,keepaspectratio]{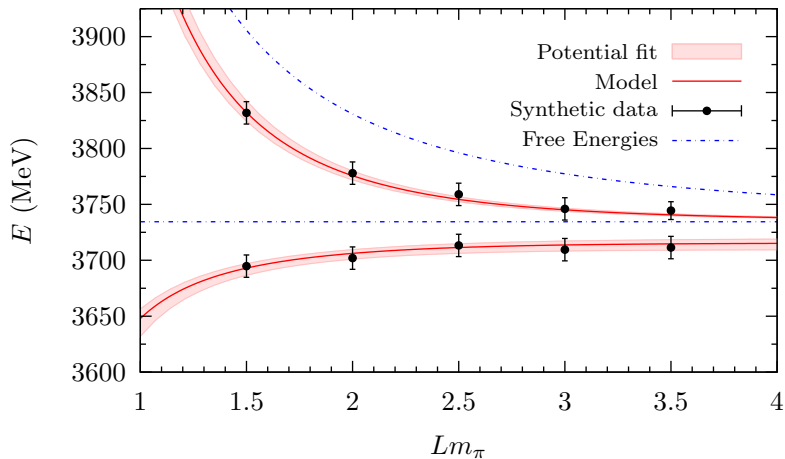}
\caption{Some energy levels for the $I=0$, $J^{PC}=0^{++}$ $D\bar{D}$
  interaction as a function of the box size $L$. The levels obtained
  with the model of Refs.\cite{Nieves:2012tt,HidalgoDuque:2012pq} in a box for $\Lambda=$ 1 GeV are shown with (red)
  solid lines, while the generated levels for some particular values
  of $L$ (synthetic data points, see the text for details), together
  with their assigned errors are displayed with black circles. The non-interacting
  energies ($m_1+m_2+ (2\pi/L)^2 n^2/2\mu$ with $n=0,1$) are shown
  with (blue) dash-dotted lines. The error bands around the solid
  lines are obtained from the fit to a potential discussed in
  Subsec.~\ref{subsec:potential}. They have been obtained by
  considering pairs of fitted parameters  ($1/C_{0a},\Lambda$) 
  that provide values of $\chi^2$ that differ from the minimum one by
  less than one unit ($\chi^2 \leqslant
\chi^2_\text{min} + 1$).\label{fig:DDbar0++levels}}
\end{figure}
\begin{figure}[ht]\centering
\includegraphics[width=0.6\textwidth,keepaspectratio]{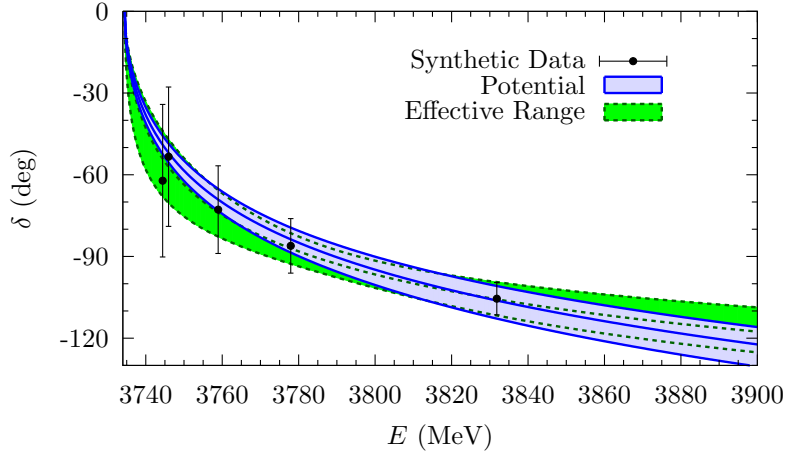}
\caption{Phase shifts obtained for $I=0$ $J^{PC}=0^{++}$ $D\bar{D}$
  interaction. The points stand for the phase shifts calculated
  from the synthetic energy levels displayed in
  Fig.~\ref{fig:DDbar0++levels} using
  Eq.~\eqref{eq:kcotd_luscher}. The green dashed line and its
  associated error band corresponds to the effective range analysis of
  Subsec.~\ref{subsec:phase}, while the blue solid line and its error band
  stand for the  results obtained by fitting a potential discussed in
  Subsec.~\ref{subsec:potential}. The phase shifts in the infinite
  volume are very similar to the latter ones, so we do not show
  them. In both cases, the error bands have been obtained by
  considering pairs of fitted parameters [($1/a, r$) for the effective
  range fit   and ($1/C_{0a},\Lambda$) for the case of the potential fit]
  that provide values of $\chi^2$ that differ from the minimum one by
  less than one unit (points included in the dark blue $\chi^2 \leqslant
\chi^2_\text{min} + 1$ ellipse displayed in  Fig.~\ref{fig:cutoff_correlation}).
\label{fig:ProblemaInverso_1}}
\end{figure}

We start discussing the case of the isoscalar $0^{++}$ $D\bar{D}$
interaction. Some levels found from  the model of
Ref.~\cite{HidalgoDuque:2012pq}  in a finite
box,\footnote{In what follows we will use an UV cutoff $\Lambda=1$ GeV
  when presenting results deduced from the model of
  Ref.~\cite{HidalgoDuque:2012pq}, both for finite boxes and in the
  infinite volume case. Other cutoffs compatible with the effective
  theory designed in \cite{Nieves:2012tt,HidalgoDuque:2012pq} give
  rise to similar results.} obtained as the zeros of
Eq.~\eqref{eq:tmat_FV}, are shown with a (red) solid line in
Fig.~\ref{fig:DDbar0++levels}. The synthetic levels generated from
them and our choice for their errors are shown with points. Recall
that we give an error of $\pm 10\ \text{MeV}$ to these points trying
to simulate a realistic situation in a LQCD study, where these
levels will be determined with some statistical uncertainties. From
the upper level and the L\"uscher's formula,
Eq.~\eqref{eq:kcotd_luscher}, we find the phase shifts shown with
points in Fig.~\ref{fig:ProblemaInverso_1}. The errors in the phase
shifts in this figure are determined by recalculating them, through
Eq.~\eqref{eq:kcotd_luscher}, with different values of the upper level
energy, $E$, randomly taken within the error intervals displayed for each of the
synthetic data points in Fig.~\ref{fig:DDbar0++levels}.

We could also obtain the scattering length and the effective range parameters
either from the determined phase shifts, or from
Eqs.~(\ref{eq:eff_range}) and (\ref{eq:kcotd_luscher}). Actually,
combining these two latter equations we have,  
\begin{equation}
\text{Re} \, \delta G_L= \lim_{\Lambda \to \infty}
\text{Re}\left(\tilde{G}(E)-G(E) \right)~= -\frac{2\pi}{\mu} 
\left(-\frac{1}{a} + \frac{1}{2}
r k^2 + \cdots\right) \label{eq:strategia}
\end{equation}
for the upper energy levels, $E$, determined in finite boxes of
different sizes. We have obtained $1/a$ and $r$ from a $\chi^2-$linear fit to
the five data points generated for $\text{Re} \, \delta G_L$ using the five synthetic
upper energy levels\footnote{To estimate the errors in $\text{Re} \,
  \delta G_L$ for each of the synthetic energy levels considered, we
  follow a procedure similar to that outlined above for the
  phase shifts. Thus, we let the synthetic energy level vary within the
  error interval displayed in Fig.~\ref{fig:DDbar0++levels} and find
  the range of variation of $\text{Re} \,
  \delta G_L$. } shown in Fig.~\ref{fig:DDbar0++levels}. We find
\begin{equation}
\frac{1}{a} = 0.62 \pm 0.25~{\rm fm}^{-1}, \qquad r=  0.53 \pm 0.18 ~{\rm fm}  \label{eq:bestfit_effr_phase}
\end{equation}
with a linear Gaussian correlation coefficient $R=0.83$. From the above
result, we find 
\begin{equation}
a = 1.6^{+1.0}_{-0.5}\ \text{fm}~.
\end{equation}
These values are to be compared with those obtained in the infinite
volume model, Eqs.~\eqref{eq:sl_th} and \eqref{eq:er_th}, with
parameter $C=C_{0a}(\Lambda =
1\ \text{GeV}) = -1.024\ \text{fm}^2$, which turn out to be:
\begin{equation}
a_\text{th} = 1.38\ \text{fm}~,           \quad r_\text{th} = 0.52\ \text{fm}~. \label{eq:scatteringlength_0++_theory}
\end{equation}
Our fitted values are compatible with the theoretical ones, but have
sizeable errors, although the correlation is large. Performing a standard
analytical continuation of Eqs.~\eqref{eq:tmat_kcotd} and \eqref{eq:eff_range} 
 below the $D\bar{D}$ threshold, we estimate the position of 
the $X(3715)$ bound state, 
\begin{equation}
E=3721^{+10}_{-25}\ \text{MeV}~, \label{eq:bind1}
\end{equation}
whereas the value found in Ref.~\cite{HidalgoDuque:2012pq} is
$E=3715^{+12}_{-15}\ \text{MeV}$.\footnote{The errors calculated for
  finite volume quantities in this work refer to the statistical
  uncertainties we generate in the synthetic data. The errors quoted
  from Ref.~\cite{HidalgoDuque:2012pq} refer
  instead to the uncertainties in the determination of the constants
  appearing in the potential.} The binding energy, $B<0$, is obtained from
Eqs.~\eqref{eq:tmat_kcotd} and \eqref{eq:eff_range}, upon changing $k
\to i \kappa$, and imposing $T^{-1}=0$, 
\begin{equation}\label{eq:bindingenergy_from_EffR}
B = \frac{\kappa^2}{2\mu}~, \qquad \kappa = \frac{1 \pm \sqrt{1-2r/a}}{r}~.
\end{equation}
To estimate the
uncertainties  in Eq.~(\ref{eq:bind1}), we have performed a Monte Carlo
simulation taking into account the existing
statistical correlations between $1/a$ and $r$. We quote a 68\% confident
interval (CL), but with some caveats as we explain next. Note that $2r/a$
is not far from unity and within errors it can be even bigger, which
means that we can get some events in the Monte Carlo runs (around 25\%)
with $1-2r/a <0$, for which we set the square root to zero. Thus, the
lower error quoted in Eq.~(\ref{eq:bind1}) is somehow uncertain,
since the above procedure tends to accumulate events around 3695
MeV. On the other hand, for the cases with $1-2r/a>0$, but small, the
two roots of $\kappa$ in Eq.~\eqref{eq:bindingenergy_from_EffR} are
not so different, and hence there is some ambiguity in the binding
energy $B$ (we choose the smallest value of $\kappa$). Note that,
although the value of $E$ obtained with its errors seems quite
accurate, when one considers it relative to the binding energy $B$,
we find a large dispersion, since the $D\bar{D}$ threshold is at
around $3734\ \text{MeV}$.

Finally, if we decrease the error of the synthetic energy levels from
$10\ \text{MeV}$ to $5\ \text{MeV}$, then the errors of the phase shifts as
well as those of the threshold parameters are also reduced
approximately to half of their
previous values, and the predicted mass is more accurate,
$E=3723^{+5}_{-11}\ \text{MeV}$ (and now only for around 
6\% of the Monte Carlo events, $1-2r/a$ become negative). 
This should give an idea of the precision
needed in the determination of the energy levels in order to have an
appropriate determination of the mass. 

In the next sections 
we discuss different alternatives that allow to achieve a better precision.

\subsection{Inverse analysis: fit to a potential}
\label{subsec:potential}
\begin{figure}[ht]\centering
\includegraphics[width=0.6\textwidth,keepaspectratio]{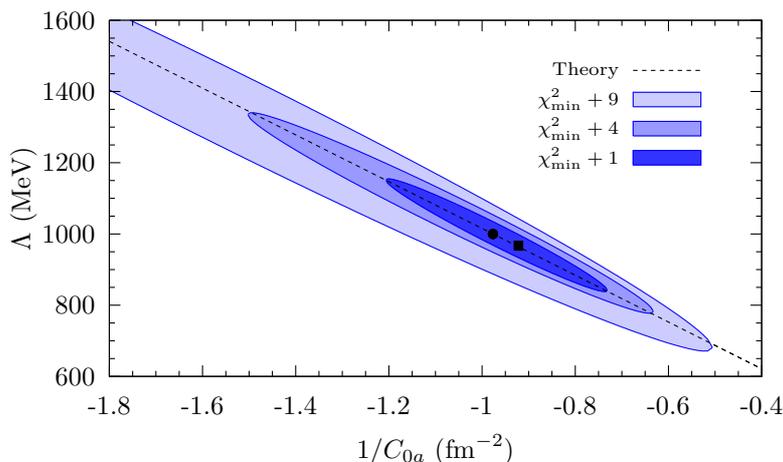}
\caption{Contour plots for the $\chi^2$ function defined in Eq.~\eqref{eq:chi2}
  for the $I=0$ $J^{PC}=0^{++}$ $D\bar{D}$ channel. The
  dashed line shows the correlation predicted from the model of 
  Ref.~\cite{HidalgoDuque:2012pq},
  Eq.~\eqref{eq:c1_GR}. The circle represents the central value taken
  for that model, $C_{0a}=-1.024\ \text{fm}^2$ and
  $\Lambda=1\ \text{GeV}$, while the square stands for the results
  of the  best fit of Eq.~\eqref{eq:besfit}.\label{fig:cutoff_correlation}}
\end{figure}

We now consider another approach to analyze/use the synthetic
levels that we generated in the previous section. Here again we  aim  to falsifying
real data obtained from LQCD Monte Carlo simulations for various
finite volumes. The analysis of
phase shifts in the previous subsection necessarily takes into account
only the level above threshold in Fig.~\ref{fig:DDbar0++levels}. It
is then convenient to develop an approach that could simultaneously make use of all available levels. Thus, we propose to describe all levels using a
potential, Eqs.~\eqref{eq:potential_gr} and \eqref{eq:tmat_FV}, fitting its
parameters for such purpose. We adopt here an approach
where we fit a counter-term $C=C_{0a}$ defining the potential\footnote{We
  follow here the notation of \cite{Nieves:2012tt,HidalgoDuque:2012pq}
    where the counter-term that appears in this channel is called
    $C_{0a}$ (see Appendix \ref{app:potentials_constants}).} and the UV cutoff $\Lambda$
(involved  in the finite box loop function and in the potential (see
Eq.~\eqref{eq:potential_gr}) to the synthetic energy levels shown in
Fig.~\ref{fig:DDbar0++levels}. Thus, the $\chi^2$
function is then given by:
\begin{equation}
\chi^2 = \sum_{i=1}^{5}
\frac{\left(E^{(0)}_{\rm th V}(L_i)-E^{(0)}_i\right)^2}{\left(\Delta
  E^{(0)}_i\right)^2} + \sum_{i=1}^{5}
\frac{\left(E_{\rm th V}^{(1)}(L_i)-E^{(1)}_i\right)^2}{\left (\Delta E^{(1)}_i\right)^2}~,
\label{eq:chi2}
\end{equation}
where $E^{(0,1)}_{\rm th V}(L_i)$ are the first two energy levels
calculated from
the HQSS potential, with parameters $C_{0a}$ and $\Lambda$, in a
finite box of size  $L_i$. On the other hand,
$E^{(0,1)}_i$ and  $\Delta E^{(0,1)}_i$ are the synthetic levels, that
we have generated, together with their assigned errors. Here, the superscript
$j=0,1$ refers to the two levels shown in
Fig.~\ref{fig:DDbar0++levels}. The fit parameters, $1/C_{0a}$
and $\Lambda$, obtained in the best fit are
\begin{equation}
\frac{1}{C_{0a}} =-0.93  \pm 0.20~\text{fm}^{-2}, \qquad \Lambda= 970 \pm 130~ \text{MeV}~,\label{eq:besfit}
\end{equation}
with a linear Gaussian correlation coefficient $R=-0.98$. These errors, and the correlation coefficient, are calculated from the hessian of $\chi^2$ at the minimum. However, since the fit is not linear, these errors are slightly different from those obtained requiring $\chi^2 \leqslant \chi^2_\text{min} + 1$. This latter requirement gives the following non-symmetrical errors:
\begin{equation}
\frac{1}{C_{0a}} =-0.93^{+0.18}_{-0.27}~\text{fm}^{-2}, \qquad \Lambda= 970^{+180}_{-120}~\text{MeV}~,\label{eq:besfit_asym}~.
\end{equation}
From Eq.~\eqref{eq:besfit}, we find $C_{0a}=-1.08^{+0.19}_{-0.29}\ \text{fm}^2$. The central values
of both, the counter-term and the UV cutoff, agree well with those of
the original model of Ref.~\cite{HidalgoDuque:2012pq}, $C_{0a} =
-1.024\ \text{fm}^2$ and $\Lambda = 1\ \text{GeV}$, used to generate
the synthetic levels. However, as expected, the two parameters are
strongly correlated. This is further discussed in Appendix~\ref{app:parameters}. A contour plot of the $\chi^2$ function in the $(1/C_{0a},\Lambda)-$plane is shown in
Fig.~\ref{fig:cutoff_correlation}, that manifestly shows the
correlation. 

On the other hand, the fitted parameters of Eq.~\eqref{eq:besfit} predict
a value for the mass of the bound state of
$E=3715^{+3}_{-6}\ \text{MeV}$ (68\% CL, obtained from a Monte Carlo Gaussian simulation keeping the statistical correlations)
in the infinite volume case. The central value agrees remarkably well
with the value obtained from the model of
Ref.~\cite{HidalgoDuque:2012pq}, $E=3715\ \text{MeV}$, and certainly
much better than that obtained with the phase shift analysis carried
out in the previous subsection ($E=3721^{+10}_{-25}\ \text{MeV}$). The
errors found now are also significantly smaller.

Finally, we have calculated error bands for the predicted finite box
 levels and phase shifts by
 quantifying the variations that are produced in these observables
 when  one randomly considers pairs ($1/C_{0a},\Lambda$)
 of parameters   that provide $\chi^2 \leqslant
\chi^2_\text{min} + 1$ (points included in the dark blue 
ellipse displayed in  Fig.~\ref{fig:cutoff_correlation}). These error bands are shown in
Figs.~\ref{fig:DDbar0++levels} and \ref{fig:ProblemaInverso_1}, respectively.

\subsection{Inverse analysis: effective range}
\label{subsec:neweffectiverange}
We have seen in Subsecs.~\ref{subsec:phase} and \ref{subsec:potential} that the fit of the synthetic energy levels with a potential leads to better results for the mass of the bound state than those obtained from the fit to the phase shifts (deduced from the upper level) with the effective range expansion. We believe that the reasons for this improvement are basically three. First, the potential fit takes into account both levels, above and below threshold, while the phase shifts analysis takes into account just the upper level. Second, in the potential fit, the ``observables'' are the energy levels, while, in the phase shifts analysis, the quantity that enters in the $\chi^2$ function is $k \cot \delta$, and the propagation of errors can lead thus to worse determinations of the parameters. Third, the analytical structure of the inverse of the amplitude is different in both approaches. In the effective range approach, ones truncates a series up to $k^2$, while in the potential fit one is effectively including further terms beyond the latter ones. Indeed, the full loop function $G$ is taken into account. To study the importance of the first two points, we follow here another approach, in which we shall keep the effective range approximation for the amplitude, but fit the energy levels (above and below threshold) instead of the phase shifts obtained from the above threshold level.

\begin{figure}[t]\centering
\includegraphics[height=6cm,keepaspectratio]{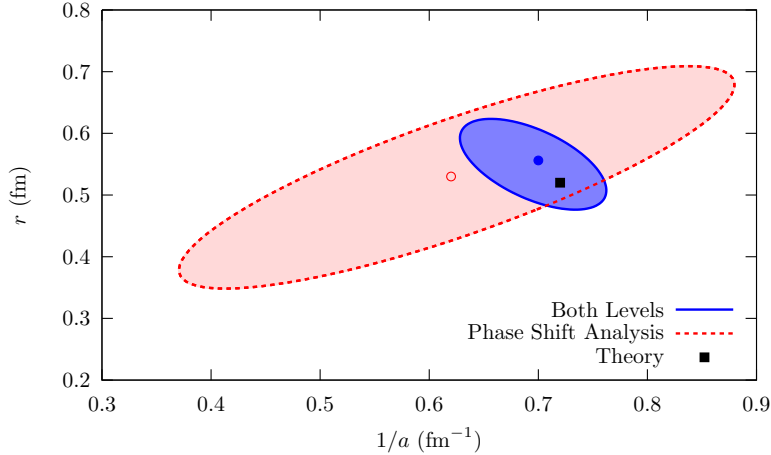}
\caption{Comparison of the determination of the effective range parameters, $1/a$ and $r$, with the methods explained in Subsecs.~\ref{subsec:phase} (red dashed line) and \ref{subsec:neweffectiverange} (solid blue line). The ellipses are obtained from the condition $\chi^2 \leqslant \chi_\text{min}^2 + 1$ in each case. The central values of each fit are represented with points. We also show, for comparison, the theoretical values of the parameters, as given in Eq.~\eqref{eq:scatteringlength_0++_theory}, with a black square.\label{fig:comparison_ERE}}
\end{figure}

The effective range expansion for the inverse of the $T$-matrix is written from Eqs.~\eqref{eq:tmat_kcotd} and \eqref{eq:eff_range}:
\begin{equation}\label{eq:neweffrange}
T^{-1}(E) = \left\{ \begin{array}{lr}
- \frac{\mu}{2\pi} \left( - \frac{1}{a} + \frac{r}{2} k^2 - i k \right)~, & \quad k=\sqrt{2\mu (E - m_1 - m_2)}~,\ E > m_1 + m_2 \\
- \frac{\mu}{2\pi} \left( - \frac{1}{a} - \frac{r}{2} \gamma^2 + \gamma \right)~, & \quad \gamma=\sqrt{2\mu (m_1 + m_2 - E)}~,\ E < m_1 + m_2
\end{array}
\right.~.
\end{equation}
Now, the energy levels in the box are found for given values of $a$ and $r$, by means of Eq.~\eqref{eq:euluscher}. It is to say, by numerically solving $T^{-1}(E) = \delta G_L (E)$, similarly as it is done in the case of the potential, but now using Eq.~\eqref{eq:neweffrange} to model the $T-$matrix both above and below threshold. We will denote the levels obtained in this manner  as 
$E^{(j)}_{\rm thEF}$. To determine the best values of $a$ and $r$, we consider thus a $\chi^2$ function as Eq.~\eqref{eq:chi2}, where the $E_i^{(j)}$ are still the synthetic levels we have generated, but replacing the $E_\text{thV}^{(j)}$ by $E^{(j)}_{\rm thEF}$, calculated as explained above. The values of the best parameters are:
\begin{equation}
\frac{1}{a} = 0.70 \pm 0.07~{\rm fm}^{-1}, \qquad r=  0.56\pm 0.07 ~{\rm fm}~, \label{eq:bestfit_effr_levels}
\end{equation}
with a linear Gaussian correlation coefficient $R=-0.6$.\footnote{In this case, the errors calculated from the requirement $\chi^2 \leqslant \chi^2_\text{min} + 1$ almost coincided with those given here in Eq.~\eqref{eq:bestfit_effr_levels}.} Hence, we obtain:
\begin{equation}
a = 1.43^{+0.16}_{-0.13}\ \text{fm}~.
\end{equation}
The errors calculated in this way are clearly smaller than those displayed in Eq.~\eqref{eq:bestfit_effr_phase} with the phase shifts analysis carried out in Subsec.~\ref{subsec:phase}. Also, the central value of the scattering length agrees better with the theoretical one, Eq.~\eqref{eq:scatteringlength_0++_theory}. These improvements have a clear impact in the determination of the mass of the bound state, which is now $E_B = 3716^{+4}_{-5}\ \text{MeV}$ (68\% CL), with smaller errors and better central value than those obtained with the phase shifts analysis in Eq.~\eqref{eq:bind1}. In Fig.~\ref{fig:comparison_ERE} we show a comparison of the ellipses in the $(1/a,r)$ parameter space determined by the condition $\chi^2 \leqslant \chi_\text{min}^2 + 1$ for the fits of Eqs.~\eqref{eq:bestfit_effr_phase} and \eqref{eq:bestfit_effr_levels}. There, it can be clearly appreciated the significant improvement achieved by fitting directly to both, lower and upper energy levels, instead of fitting to the phase shifts deduced from the latter levels. Finally, we must point out that the determination of the energy levels obtained with this method are very similar to those obtained in Subsec.~\ref{subsec:potential} by introducing of a potential, and shown in Fig.~\ref{fig:DDbar0++levels}. Actually, the differences between the upper and lower energy level curves  (and their error bands) deduced from both methods would not be easily appreciated in Fig.~\ref{fig:DDbar0++levels}. For this reason, we have not shown in this figure the results obtained from the method discussed in this subsection.

\subsection{Inverse analysis: bound state fit} 
\label{subsec:bound_state_fit}
We have discussed in Subsec.~\ref{subsec:results_model_box} the
volume dependence of the sub-threshold levels that arise when we put
the model in a finite box. For cases with $V < 0$, the potential is
attractive, and hence, a bound state in the infinite volume case may
arise. Whether it is bound or not in the infinite volume case, there would
appear a sub-threshold level for finite volumes. It was argued in Subsec.~\ref{subsec:results_model_box} 
that it may be not very clear, at first sight, if these levels tend to
the threshold energy or to a bound state in the  $L\to\infty$ limit. To
circumvent this problem, we suggest here a method to study this volume
dependence. By subtracting Eqs.~\eqref{eq:tmat_ND} and
\eqref{eq:tmat_FV}, we can write the amplitude in the finite box as
\cite{nonogo}:
\begin{equation}
\tilde{T}^{-1} = T^{-1} - \delta G_L, \qquad \delta G_L=\lim_{\Lambda\to
  \infty}\delta G = 
\lim_{\Lambda\to \infty}\left(\tilde{G} - G\right)~.
\end{equation}
A bound state with mass $E_B$ appears as a pole in the $T$-matrix,
thus in the vicinity of the pole, we can approximate:
\begin{equation}
T(E \simeq E_B) = \frac{g^2}{E-E_B} + \ldots~,\label{eq:boundapprox}
\end{equation}
where the ellipsis denote regular terms in the Laurent series of the
amplitude. The coupling can also be calculated analytically,
\begin{equation}
g^2 = \lim_{E\to E_B} (E-E_B) T(E) \quad \text{or} \quad \frac{1}{g^2} = \left. \frac{\mathrm{d} T^{-1}(E)}{\mathrm{d} E} \right\rvert_{E=E_B}~.
\end{equation}
The volume dependence of the sub-threshold level in the finite box,
given by the equation $\tilde{T}^{-1}(E) = 0$ is then dominated by this bound state,
and hence:
\begin{equation}
\tilde{T}^{-1} (E) \simeq \frac{E-E_B}{g^2} - \delta G_L(E) = 0~,
\end{equation}
from where one can write:
\begin{equation}\label{eq:bound_state_L}
E(L) = E_B + g^2 \delta G_L\left[E(L),L\right]~.
\end{equation}
This equation is a reformulation of a similar result obtained in
Refs.~\cite{Luscher:1985dn, Beane:2003da, Beane:2011iw}.\footnote{In the same line as in those references, but using boosted reference frames, in Ref.~\cite{davoudi} linear combinations of energy levels are suggested to reduce the volume dependence. Our method does not rely upon the analytical form of the volume dependence to make extrapolations since for every $L$ considered, the exact $L$ dependence is provided by $\delta G_L (E(L),L)$.} The coupling $g$
obtained here is related to $Z_\psi$ of
Ref.~\cite{Beane:2011iw}. Note, however, that
Eq.~\eqref{eq:bound_state_L} is appropriate as long as
Eq.~\eqref{eq:boundapprox} is sufficiently accurate to describe the
infinite volume $T-$matrix for the energy levels found in the lattice
simulation (i.e., energies for which $\tilde{T}^{-1}$
vanishes). Hence, the larger the box sizes, the better
Eq.~\eqref{eq:bound_state_L} will perform.\footnote{On the other hand, 
for very small binding energies, some subtleties appear, 
because the coupling $g^2$
tends to zero as the mass of the bound state approaches the
threshold~\cite{Toki:2007ab, Albaladejo:2012te, danijuan}. We will discuss this issue at length in
Subsec.~\ref{subsec:2plusplus}.}  We
extract the mass and the coupling of the bound state from a fit to the
sub-threshold level in Fig.~\ref{fig:DDbar0++levels} with the
following $\chi^2$ function,
\begin{equation}\label{eq:chi2_BS}
\chi^2 = \sum_{i=1}^5\frac{\left(E(L_i)-E^{(0)}_i\right)^2}{\left(\Delta
  E^{(0)}_i\right)^2}
\end{equation}
where $E(L)$ is given by Eq.~\eqref{eq:bound_state_L}.\footnote{It is
worth noting the following technical detail. In principle, $E(L)$
should be extracted for each $L_i$ as the implicit solution in
Eq.~\eqref{eq:bound_state_L} for given $E_B$ and $g^2$. For practical
purposes, though, it is more convenient to obtain $E(L)$ by plugging
into the right-hand side of this equation the values of $E_i^{(0)}$
and $L_i$ that we are fitting to. If Eq.~\eqref{eq:boundapprox} is
accurate enough, both methods are equivalent, as long as the effects in
Eq.~\eqref{eq:bound_state_L} of the statistical fluctuations of the
measured lattice levels are sufficiently small. In that case, the
results for $E_B$ and $g^2$ should not be very different, as we have
checked. Indeed, the best fit results  given in
Eq.~\eqref{besfitresultsg2} have been obtained within this
approximation. However, this approximation cannot be safely used when
the bound state is placed close to threshold, because then $\delta
G_L$ rapidly changes and statistical fluctuations in the determined
lattice energy levels induce large variations in the right-hand side of
Eq.~\eqref{eq:bound_state_L}.} The best $\chi^2$ is
  $\chi^2_\text{min} = 0.5$, and the parameters so obtained are:
\begin{equation}
E_B = 3712 \pm 6\ \text{MeV}~, \qquad g^2 = \left( 2.8 \pm 2.1
\right)\ \text{GeV}^{-1}~, \label{besfitresultsg2}
\end{equation}
to be compared with those obtained with the model in the infinite
volume case, $E_B = 3715\ \text{MeV}$ and $g^2 =
2.6\ \text{GeV}^{-1}$.

It could be that, for the case of weakly bound states, the error bars
on the energies overlap with the threshold and it is difficult to
determine if one has a bound state or not. A weak attractive potential
that does not bind in the infinite volume case, still provides a
level below threshold for finite volumes, and the energies go to
threshold as $L\to\infty$. For this case we proceed as follows. The
volume dependence of this level would be given by:
\begin{align*}
\tilde{T}^{-1} = 0 & = T^{-1} - \delta G_L = -\frac{\mu}{2\pi}
\left(-\frac{1}{a} 
+ \frac{r}{2}k^2  + \cdots \right) - \frac{2\mu}{L^3}\frac{1}{k^2} - \alpha - \beta k^2 + \cdots~\\
& = -\frac{2\mu}{L^3 k^2} + \frac{\mu}{2\pi a}- \alpha 
+ \left( -\frac{\mu r}{4\pi}-\beta \right) k^2 + \cdots 
\end{align*}
with some coefficients $\alpha$ and $\beta$, disregarding
exponentially suppressed terms. In the above equation, we have
explicitly separated the threshold singularity of $\delta G_L$,
\begin{equation*}
\frac{1}{L^3} \frac{1}{E-m_1-m_2} = \frac{2\mu}{L^3 k^2}~.
\end{equation*}
Hence, the general behavior of this level would be:
\begin{equation}
k^2 = \frac{2\mu}{L^3} \frac{1}{A + B k^2}~.
\end{equation}
Now, this expression could be used in a $\chi^2$ function as in
Eq.~\eqref{eq:chi2_BS} (with $k^2 < 0$ for the level below threshold).
We have checked that, if we try to fit the energies of the lower level
of Fig.~\ref{fig:DDbar0++levels} with this formula, we get a much
worse $\chi^2$ value, discarding the possibility that there is not a
bound state in the infinite volume.

We have seen then that the method outlined in this subsection allows
for a safe discrimination between those levels that correspond to
bound states in the infinite volume and those that do not, and it also
gives a precise determination of the mass. It is also worth noting
that the errors for the mass of the bound state are similar to those
obtained with the analysis from a potential in
Subsec.~\ref{subsec:potential}, and smaller than those calculated
with the phase shift analysis in Subsec.~\ref{subsec:phase}.

\subsection{\boldmath The case of the $I=0$ $J^{PC}=2^{++}$ channel} 
\label{subsec:2plusplus}

\begin{figure}[t]\centering
\includegraphics[width=0.6\textwidth,keepaspectratio]{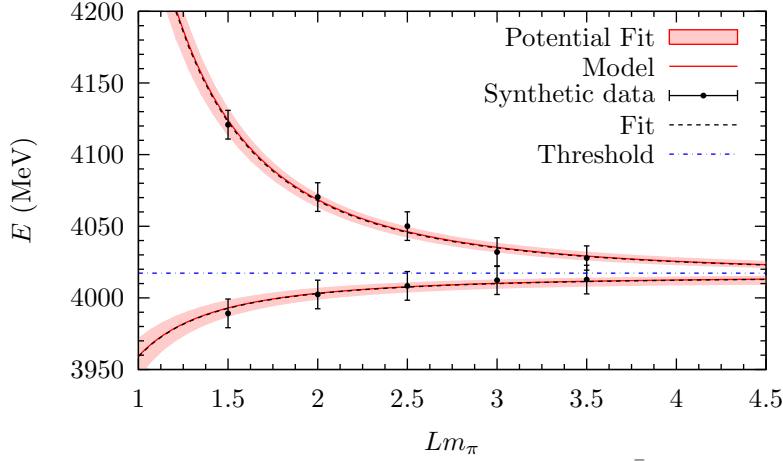}
\caption{Same as in Fig.~\ref{fig:DDbar0++levels}, but for the $I=0$ $2^{++}$ $D^\ast\bar{D}^\ast$
  interaction.\label{fig:2plusplus_levels}}
\end{figure}

\begin{figure}[t]\centering
\includegraphics[width=0.6\textwidth,keepaspectratio]{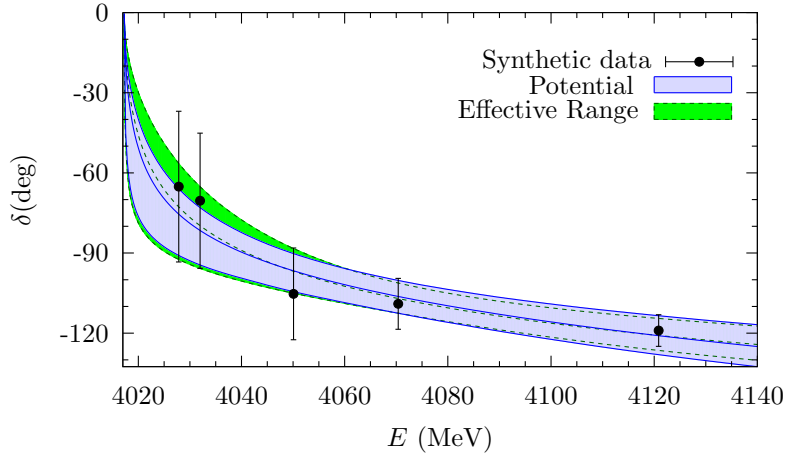}
\caption{Phase shifts for the $I=0$ $J^{PC}=2^{++}$ $D^\ast \bar{D}^\ast$ channel interaction. The points stand for the synthetic phase shifts generated from the upper energy level of Fig.~\ref{fig:2plusplus_levels}. The (green) dashed line correspond to the effective range fit, while the (blue) solid line correspond to the potential fit. The associated error bands are obtained by considering randomly chosen pair of parameters [$(1/a,r)$ and $(1/C_0,\Lambda)$ for the effective range and potential fits, respectively] that satisfy $\chi^2 \leqslant \chi^2_{\text{min}} + 1$. \label{fig:2plusplus_phases}}
\end{figure}

We now repeat the same analyses carried out above but now for the case of the bound state present in the $I=0$ $J^{PC}=2^{++}$ $D^\ast \bar{D}^\ast$ channel. As already mentioned, the $2^{++}$ state is a HQSS partner of the $X(3872)$ molecule which dynamics, at LO in the heavy quark expansion, is being
determined by precisely the same combination of counter-terms that appear in the $X(3872)$ channel. The existence of the $2^{++}$ state, either as a bound state or a resonance, is a quite robust consequence of HQSS \cite{HidalgoDuque:2012pq, Nieves:2012tt}, and it can be certainly subject to experimental detection. It is also worth to discuss this channel in this context because, contrary to the case analyzed before, we have here a very weakly bound state. In the calculation of Ref.~\cite{HidalgoDuque:2012pq}, and also in the results of Subsec.~\ref{subsec:results_model_box}, shown in the left panel of Fig.~\ref{fig:BS_from_model_1}, charged and neutral coupled channels were studied, because the gap between both thresholds is indeed larger than the binding energy of the bound state. The mass of the bound state is $E_B = 4013.2\ \text{MeV}$, whereas the neutral and charged thresholds are located at $4014.0\ \text{MeV}$ and $4020.6\ \text{MeV}$, respectively. Here, however, in order to discuss the problem in simpler terms, we will consider only an uncoupled channel problem with $I=0$, and use isospin average masses, keeping the relevant counter-term, $C_0$ in the nomenclature of Ref.~\cite{HidalgoDuque:2012pq}, to the same value, namely, $C_0 = -0.731\ \text{fm}^2$. In this way, the threshold is located at $4017.3\ \text{MeV}$, whereas the mass of the bound state now becomes $E_B = 4014.6\ \text{MeV}$. The first two energy levels obtained with this simplified model are shown with (red) solid lines in Fig.~\ref{fig:2plusplus_levels}. As before, for the following statistical analyses we consider the synthetic levels shown in this figure with points. The centroid of these points is randomly shifted in the range $\pm 5\ \text{MeV}$, and they are given an error of $\pm 10\ \text{MeV}$. With the points of the upper level, we generate the phase shifts shown in Fig.~\ref{fig:2plusplus_phases} with points, through Eq.~\eqref{eq:kcotd_luscher}. We can now obtain the scattering length and effective range as in Subsec.~\ref{subsec:phase}, that turn out to be:
\begin{equation} \label{eq:phase_2++}
\frac{1}{a} = 0.41 \pm 0.30\ \text{fm}^{-1}~,\quad r=0.67 \pm 0.19\ \text{fm}~,
\end{equation}
with a linear Gaussian correlation $R = 0.81$. From the above fitted value for $1/a$, we find:
\begin{equation}
a = 2.4^{+2.4}_{-1.2}\ \text{fm}\ (68\%\ \text{CL})~,
\end{equation}
while the theoretical values, obtained from Eqs.\eqref{eq:sl_th} and \eqref{eq:er_th}, are:
\begin{equation} \label{eq:theory_2++}
a_{\text{th}} = 3.0\ \text{fm}~,\quad r_{\text{th}} = 0.58\ \text{fm}~,
\end{equation}
which agree with the above determinations, within errors. The phase shifts obtained with these parameters, and the associated error bands, are shown in Fig.~\ref{fig:2plusplus_phases} with (green) dashed lines, and they satisfactorily reproduce the synthetic data. The mass of the bound state turns out to be then $E_B = 4013^{\hphantom{1}+4}_{-18}\ \text{MeV}$. Recall that the caveats raised in Subsec.~\ref{subsec:phase} apply here.


The next step is to consider a potential fit, as in Subsec.~\ref{subsec:potential}, where the free parameters are $C_0$ and $\Lambda$. From a best fit, as described in Subsec.~\ref{subsec:potential}, we find:
\begin{equation}
\frac{1}{C_0} = -1.4\pm 0.5\ \text{fm}^{-2}~, \quad \Lambda = 1020\pm 240\ \text{MeV}~,\label{eq:bestfit_2++_pot}
\end{equation}
with a linear Gaussian correlation coefficient $R=-0.97$. The non-symmetrical errors given by the condition $\chi^2 \leqslant \chi^2_\text{min} + 1$ are instead:
\begin{equation}
\frac{1}{C_0} = -1.4^{+0.4}_{-0.7}\ \text{fm}^{-2}~, \quad \Lambda = 1020^{+360}_{-190}\ \text{MeV}~.
\end{equation}
The energy levels obtained with these parameters are shown in Fig.~\ref{fig:2plusplus_levels} with a (black) dashed line, although they are so similar to those of the exact model that they mostly overlap. Also shown in the figure are the error bands generated from the errors of the parameters, as described previously in Fig.~\ref{fig:DDbar0++levels}. With this potential, we can also calculate the phase shifts, which are shown in Fig.~\ref{fig:2plusplus_phases} with (blue) solid lines, and also the associated error band. The quality of both descriptions, that of the effective range and that of the potential, are very similar, and indeed both lines are very similar to the one of the exact model, and hence we do not show the latter. The value of $C_0$ deduced from Eq.~\eqref{eq:bestfit_2++_pot} is $C_0 = -0.71^{+0.19}_{-0.39}\ \text{fm}^2$ (68\% CL), in good agreement with the one of the infinite volume model, $C_0 = -0.73$. Finally, the mass of the bound state is given by $E_B = 4014.3^{+2.3}_{-5.4}\ \text{MeV}$. We must stress again that the errors obtained with this method are smaller than those obtained with the phase shifts analysis.

Now, we consider the method of Subsec.~\ref{subsec:neweffectiverange}, in which the effective range expansion is used to study the levels below and above threshold. In this case, the best fit values that we obtain for the scattering length and effective range parameters are:
\begin{equation}
\frac{1}{a} = 0.35\pm 0.15\ \text{fm}^{-1}~,\qquad r=0.64\pm 0.15\ \text{fm}~,\label{eq:bestfit_2++_ere}
\end{equation}
with a linear Gaussian correlation coefficient $R=0.3$. The non-symmetric errors stemming from the condition $\chi^2 \leqslant \chi^2_\text{min} + 1$ turn out to be:
\begin{equation}
\frac{1}{a} = 0.35^{+0.13}_{-0.21}\ \text{fm}^{-1}~,\qquad r=0.64^{+0.14}_{-0.16}\ \text{fm}~.
\end{equation}

Propagating the correlated Gaussian errors of Eq.~\eqref{eq:bestfit_2++_ere}, we find:
\begin{equation}
a = 2.9^{+2.0}_{-0.9}\ \text{fm}~.
\end{equation}
We note that here, as it also occurred for the $I=0$, $J^{PC}=2^{++}$ case, the central values obtained with this method agree better with the theoretical ones, Eq.~\eqref{eq:theory_2++}, than those obtained with the phase shift description, Eq.~\eqref{eq:phase_2++}, and have smaller errors than the latter ones. The mass of the bound state obtained is $E_B = 4014.2^{+2.3}_{-4.8}\ \text{MeV}$, which is better determined than that obtained by means of the phase shifts analysis, and very similar to the one obtained with the potential method.

Finally, we should proceed now with the analysis performed in Subsec.~\ref{subsec:bound_state_fit}. However, there is a major difference in this case, namely, that the bound state is very close to the threshold in this case. It is known that, in this case, the coupling of the state tends to zero \cite{Toki:2007ab, Albaladejo:2012te, danijuan}, and so additional terms in the Laurent series, Eq.~\eqref{eq:boundapprox}, are relevant for energies not very far from the bound state mass.\footnote{As an example, consider a background term in the amplitude in Eq.~\eqref{eq:boundapprox}, so that $T = g^2/(E-E_B) + \beta + \cdots$. From the theoretical model, one can calculate $g^2 = 0.58\cdot 10^{-3}\ \text{MeV}^ {-1}$ and $\beta = -0.68\cdot 10^{-4}\ \text{MeV}^{-2}$. For energies $E \simeq 3990\ \text{MeV}$, as we find for the lower level in Fig.~\ref{fig:2plusplus_levels} for $Lm_\pi = 1.5$, we have $\left\lvert g^2/(E-E_B) \right\rvert < \left\lvert \beta \right\rvert$.} Further, since we are considering an error of $\pm 10\ \text{MeV}$ in the energy levels, and we are trying to reproduce a bound state of binding energy of $2-3\ \text{MeV}$, we should expect a greater qualitative impact on the lower energy levels, which are the only ones considered in this method (note that the other methods examined above use always the upper levels as well). These considerations explain why, when performing such analysis, we obtain very bad results for the bound state mass and the coupling. Hence, we must conclude that this method can only be applied safely in the case of bound states that are not very loosely bound. On the other hand, including a background term would increase the number of free parameters, and so, the errors stemming from a best fit would be even larger. At this stage, the approaches in Subsecs.~\ref{subsec:phase} to \ref{subsec:neweffectiverange} would be more useful for cases like this one, in which the bound state is very close to threshold.

\section{Conclusions}

%{\bf Some relevant technical aspects:}
%\begin{itemize}
%\item Eq. 4.13 
%\begin{equation}
%E(L) = E_B + g^2 \delta G_L\left[E(L),L\right]~.
%\end{equation}
%
%an alternative way to get the bound state energy with the
%  bonus of obtaining the coupling as well (its a pity, that something
%  similar was already proposed in \cite{Luscher:1985dn, Beane:2003da, Beane:2011iw}
%\item B11 provides ${\cal Z}_{00}(1,\hat k^2)$  with enough accuracy in a computationally easy way.
%\end{itemize}

 In this paper we have addressed the interaction of heavy charmed mesons in the hidden charm sector where several bound states are produced using an interaction that is based on heavy quark spin symmetry. The interaction is then studied in a finite box and the levels expected from a lattice QCD calculation are evaluated for the $D \bar D$, $D^* {\bar D}^*$ states and their SU(3) partners. Then the inverse problem is faced, generating ``synthetic data'' from the levels obtained and using different procedures to obtain the relevant magnitudes in the infinite space, phase shifts and binding energies for the bound states. Particular emphasis is done in the error analysis to establish the accuracy of the different methods. We use two levels for different values of the box size $L$, one below threshold and the closest one above threshold. One strategy is to use the L\"uscher formula to get phase shifts for each energy of the level above threshold. Another strategy is to use the effective range approximation, but fitting the scattering length and effective range to both levels (above and below threshold). The two methods work, but the latter one gives better determinations of the parameters (scattering length and effective range), but also of the mass of the bound state. Yet, the method that proves most efficient\footnote{As one moves far away from threshold, any method based on the effective range approximation becomes less appropriated. The potential fit method  improves on the effective range expansion since it includes the full loop function and not only  the imaginary part fixed by unitarity. Thus, this latter method should work better when energies significantly lower or higher than the threshold are considered.} is to parameterize a potential and a regularizing UV cutoff for the meson meson loops and carry a fit to the data. Once the potential and the UV cutoff are determined one can evaluate the phase shifts and binding energies with much better precision than the one assumed in the ``synthetic data''. The UV cutoff is not needed if one considers levels of only one energy, but it appears when different energies are used in the fit, yet, we show that it is highly correlated with the potential.
   We also devoted particular attention to the case of weakly bound states, where special care must be taken.
   Finally, as a byproduct we present an efficient method to obtain the L\"uscher function, supported by an analytical study that allows one to truncate the sum by means of a Gaussian form factor and estimate the error induced by the truncation.

\section*{Acknowledgments}
This work is partly supported by the Spanish Ministerio de Econom\'\i a y Competitividad and
European FEDER funds under the contract number FIS2011-28853-C02-01 and  FIS2011-28853-C02-02, and the Generalitat
Valenciana in the program Prometeo, 2009/090. We acknowledge the support of the European
Community-Research Infrastructure Integrating Activity Study of Strongly Interacting Matter
(acronym HadronPhysics3, Grant Agreement n. 283286) under the Seventh Framework Programme of the EU.

\appendix

\section{\boldmath Potentials for the $H\bar{H}$ interaction}
\label{app:potentials_constants}
\newcommand{\vs}{$\vphantom{\displaystyle \frac{a}{x}}$}
\begin{table}[ht]\small
\begin{center}
\begin{tabular}{|c|c|c|c|}
\hline 
 \vs $J^{PC}$ & $H\bar{H}'$ & $^{2S+1}L_J$  & $V_C$ \\ \hline\hline
 \vs $0^{++}$ & $ D\bar{D}$ & $^1S_0$ & $C_{0a}$ \\ \hline
 \vs $1^{++}$ & $ D^*\bar{D}$ & $^3S_1$ & Eq.~\eqref{Acoplado1} \\
 \vs $1^{+-}$ & $ D^*\bar{D}$ & $^3S_1$ & $C_{0a} - C_{0b}$ \\ \hline
 \vs $0^{++}$ & $ D^*\bar{D}^*$ & $^1S_0$ & $C_{0a} - 2\,C_{0b}$ \\
 \vs $1^{+-}$ & $ D^*\bar{D}^*$ & $^3S_1$ & $C_{0a} - C_{0b}$ \\
 \vs $2^{++}$ & $ D^*\bar{D}^*$ & $^5S_2$ &  Eq.~\eqref{Acoplado1} \\ \hline
\end{tabular}
\begin{tabular}{|c|c|c|c|}
\hline
 \vs $J^{PC}$ & $H\bar{H}'$ & $^{2S+1}L_J$  & $V_C$\\ \hline\hline
 \vs $0^{+}$ & $ D\bar{D}_{s}$ & $^1S_0$ & $C_{1a}$\\ \hline
 \vs $1^{+}$ &  $ D_{s}^*\bar{D}$,$ D^*\bar{D}_{s}$ & $^3S_1$ & Eq.~\eqref{Acoplado2} \\
\vs & & & \\ \hline
 \vs $0^{+}$ & $ D^*\bar{D}_{s}^*$ & $^1S_0$ & $C_{1a} - 2\,C_{1b}$ \\
 \vs $1^{+}$ & $ D^*\bar{D}_{s}^*$ & $^3S_1$ & $C_{1a} - C_{1b}$ \\
 \vs $2^{+}$ & $ D^*\bar{D}_{s}^*$ & $^5S_2$ &  $C_{1a} + C_{1b}$\\ \hline
\end{tabular}
\end{center}
\caption{Potentials for the $H\bar{H}'$ interaction for the $I=0$ (left) and $I=1/2$ (right) different $J^{PC}$ channels.\label{tab:pot_1}}
\end{table}
\begin{table}[ht]\small
\begin{center}
\begin{tabular}{|c|c|c|c|}
\hline
 \vs $J^{PC}$ & $H\bar{H}'$ & $^{2S+1}L_J$  & $V_C$\\ \hline \hline
 \vs $0^{++}$ & $ D\bar{D}$ & $^1S_0$ & $C_{1a}$ \\ \hline
 \vs $1^{++}$ & $ D^*\bar{D}$ & $^3S_1$ & Eq.~\eqref{Acoplado1}\\
 \vs $1^{+-}$ & $ D^*\bar{D}$ & $^3S_1$ & $C_{1a} - C_{1b}$\\ \hline
 \vs $0^{++}$ & $ D^*\bar{D}^*$ & $^1S_0$ & $C_{1a} - 2\,C_{1b}$ \\
 \vs $1^{+-}$ & $ D^*\bar{D}^*$ & $^3S_1$ & $C_{1a} - C_{1b}$\\
 \vs $2^{++}$ & $ D^*\bar{D}^*$ & $^5S_2$ &  Eq.~\eqref{Acoplado1}\\ \hline
\end{tabular}
\begin{tabular}{|c|c|c|c|}
\hline
 \vs $J^{PC}$ & $H\bar{H}'$ & $^{2S+1}L_J$  & $V_C$\\ \hline\hline 
 \vs $0^{++}$ & $ D_{s}\bar{D}_{s}$ & $^1S_0$     & $C_{01a}$\\ \hline
 \vs $1^{++}$ & $ D_{s}^*\bar{D}_{s}$ & $^3S_1$   & $C_{01a} + C_{01b}$\\
 \vs $1^{+-}$ & $ D_{s}^*\bar{D}_{s}$ & $^3S_1$   & $C_{01a} - C_{01b}$ \\ \hline
 \vs $0^{++}$ & $ D_{s}^*\bar{D}_{s}^*$ & $^1S_0$ & $C_{01a} - 2 C_{01b}$ \\
 \vs $1^{+-}$ & $ D_{s}^*\bar{D}_{s}^*$ & $^3S_1$ & $C_{01a} - C_{01b}$\\
 \vs $2^{++}$ & $ D_{s}^*\bar{D}_{s}^*$ & $^5S_2$ & $C_{01a} + C_{01b}$\\ \hline
\end{tabular}
\end{center}
\caption{Potentials for the $H\bar{H}'$ interaction for the $I=1$ (left) and hidden strangeness sector (right) different $J^{PC}$ channels.\label{tab:pot_2}} 
\end{table}

In this Appendix, for completeness, we briefly review the formalism of the effective field theory derived in Refs.~\cite{HidalgoDuque:2012pq, Nieves:2012tt} to study charmed meson-antimeson bound states. This effective field theory incorporates SU(3)-light flavour symmetry and HQSS. In this context and at LO, the potential is given by contact terms, related to four independent LECs, namely $C_{0a}$, $C_{0b}$, $C_{1a}$ and $C_{1b}$, in the notation of Refs.~\cite{HidalgoDuque:2012pq, Nieves:2012tt}. Thus, the potentials in the different channels will be given by different linear combinations of these LECs. The fit of these LECs to four experimental data, as stated in Sec.~\ref{sec:infinite_volume}, allows one to fix the numerical value of the different counterterms. In the following, we summarize the form of the potentials for the different $I$ and $J^{PC}$ channels.

In the isoscalar channel, the only involved LECs are $C_{0a}$ and $C_{0b}$, and the way they appear in the different $J^{PC}$ channels is shown in Table~\ref{tab:pot_1} (left panel). Potentials in the $I=1/2$ and $I=1$ sector are the same, except in the channel where coupled channels must be considered as discussed below. These potentials only depend on $C_{1a}$ and $C_{1b}$. The explicit expressions for $I=1/2$ are given in Table~\ref{tab:pot_1} (right panel), whereas those for $I=1$ are shown in Table~\ref{tab:pot_2} (left panel). Finally, in the hidden strangeness sector, the four LECs appear, and the final potential is the arithmetic mean of the corresponding isoscalar and isovectorial interaction, as can be seen in Table~\ref{tab:pot_2} (right panel). Note that, for this table, we have defined $C_{01a} = \frac{1}{2}\left( C_{0a} + C_{1a} \right)$ and $C_{01b} = \frac{1}{2}\left( C_{0b} + C_{1b} \right)$.

However, coupled channels must be taken into account in two cases. The first one is that in which the mass difference of the charged and neutral channels thresholds is not negligible as compared to the binding energy of the state. The second case occurs when charge conjugation is not a good quantum number. The first scenario is important in the study of the $D\bar{D}^{*}$ system with $J^{PC} = 1^{++}$ and $D^{*}\bar{D}^{*}$ with $J^{PC} = 2^{++}$. In this case, the potential will be the interaction between the charged and neutral channels and will be given by the $2\times 2$ matrix:
\begin{eqnarray}
V_{0} =\frac{1}{2} \left( \begin{array}{cc} C_{0} + C_{1} & C_{0} - C_{1} \\ C_{0} - C_{1} & C_{0} + C_{1}\end{array} \right)~,
\label{Acoplado1}
\end{eqnarray}
being $C_{0} = C_{0a} + C_{0b}$ and $C_{1} = C_{1a} + C_{1b}$. The second scenario where coupled channels are important is because of the mixing between the $D_{s} \bar{D}^{*}$ and  $D \bar{D}_{s}^{*}$ channels. In this case, the potential is:
\begin{eqnarray}
V_{1} = \left( \begin{array}{cc} C_{1a} &- C_{1b} \\ -  C_{1b} & C_{1a}\end{array} \right)~.
\label{Acoplado2}
\end{eqnarray}

Using these combinations of counter-terms to describe the four input data, as explained in Sec.~\ref{sec:infinite_volume}, the numerical values of the LECs for the two values of the cutoff considered in this work are the following:

\begin{eqnarray}
C_{0a} = -3.366^{+0.024}_{-0.015}\,{\rm fm}^2  &\qquad& (-1.024^{+0.005}_{-0.003} \,{\rm fm}^2)  \, , \label{eq:c0a}\\
C_{0b} = +1.673^{+0.012}_{-0.008} \,{\rm fm}^2 &\qquad& (+0.293^{+0.004}_{-0.002}\,{\rm fm}^2) \, , \label{eq:c0b}\\
C_{1a} = -1.76^{+0.29}_{-0.29} \,{\rm fm}^2    &\qquad& (-0.684^{+0.064}_{-0.063} \,{\rm fm}^2) \, ,\label{eq:c1a} \\
C_{1b} = +1.68^{+0.15}_{-0.15} \,{\rm fm}^2    &\qquad& (+0.311^{+0.033}_{-0.033} \,{\rm fm}^2 )\, , \label{eq:c1b}
\end{eqnarray} 
for $\Lambda = 0.5\,{\rm GeV}$ ($1\,\rm GeV$). These are the values used through this work. The uncertainties in the above equations account for possible HQSS violations and errors in the input used to fix the counter-terms (see a detailed discussion in Ref.~\cite{HidalgoDuque:2012pq}). For simplicity in this exploratory work,  we have ignored them.

\section{Gaussian regulator and relation to L\"uscher formula}
\label{app:limit}
In this Appendix, we discuss the details of
Eq.~\eqref{eq:kcotd_luscher} within a Gaussian regularization scheme.
We also study the dependence of the function $\delta G(E)$, that
appeared in Eq.~(\ref{eq:euluscher}), on the UV cutoff $\Lambda$. For
convenience, we re-write $\delta G(E)$ as
\begin{eqnarray}
\delta G(E;\Lambda)& = & \tilde{G}(E) - G(E) \nonumber\\
& = & \overbrace{ 
\frac{1}{L^3}\sum_{\vec{q}}                         \frac{e^{-2(\vec{q}^{\,2}-k^2)/\Lambda^2}-1}{E-m_1-m_2 - \frac{\vec{q}^{\,2}}{2\mu}}
-
\int \frac{\text{d}^3 \vec{q}}{(2\pi)^3} \frac{e^{-2(\vec{q}^{\,2}-k^2)/\Lambda^2}-1}{E-m_1-m_2 - \frac{\vec{q}^{\,2}}{2\mu} + i0^+} 
}^{\delta G_A} \nonumber\\ 
& + & \underbrace{
\frac{1}{L^3}\sum_{\vec{q}} \frac{1}{E-m_1-m_2 - \frac{\vec{q}^{\,2}}{2\mu}}
- 
\int \frac{\text{d}^3 \vec{q}}{(2\pi)^3} \frac{1}{E-m_1-m_2 - \frac{\vec{q}^{\,2}}{2\mu} + i0^+} 
}_{\delta G_L}
\end{eqnarray}
The function $\delta G$ explicitly depends on the cutoff $\Lambda$,
and this dependence is carried by the  $\delta G_A$ term. On the other
hand, the term $\delta G_L$ is well defined, and it is related to the
L\"uscher function~\cite{Doring:2011vk} (see discussion below). In the strict  $\Lambda \to
\infty$ limit, only the second term survives, which justifies our approach in
Sec.~\ref{sec:FV}. Still, for practical purposes, the limit
$\Lambda \to \infty$ can only be achieved by taking $\Lambda$ large
enough, and then it is necessary to study the dependence of $\delta G$
with $\Lambda$. Let us note that $\delta G_A$ has no poles, and hence
it is exponentially suppressed with $L$ according to the regular
summation theorem \cite{Luscher:1986pf,Luscher:1990ux}. For $k^2>0$,
$E>m_1+m_2$, $\delta G_L$ is not exponentially suppressed for $L\to
\infty$ and, in this case, $\delta G_L$ clearly dominates over $\delta
G_A$. 

However for $k^2 < 0$, $\delta G_L$ is also exponentially suppressed
as $L$ increases, and therefore one needs to explicitly calculate the
dependence of $\delta G_A$ on $\Lambda L$.

Let us calculate the derivative of $\delta G$ with respect to
$\Lambda$. Only $\delta G_A$ depends on $\Lambda$, and
this latter function does it through the exponential function $\exp
[-2(\vec{q}^{\,2}-k^2)/\Lambda^2]$. The derivative brings down a
factor  $(\vec{q}^{\,2}-k^2)$ that cancels out the denominators. This
greatly simplifies the calculation of both  the sum and   the
integral. The latter one is trivial and it only amounts to the
integration of a Gaussian function, while the former one, up to constant
factors, now reads
\begin{equation}
\frac{1}{L^3}\sum_{\vec{q}} e^{-2\vec{q}^{\,2}/\Lambda^2} = \left(
\frac{1}{L} \sum_{n=-\infty}^{+\infty} e^{-2
  \left(\frac{2\pi}{L}\right)^2 \frac{n^2}{\Lambda^2}} \right)^3= 
\left[  \frac{\theta_3(0,e^{-\frac{8\pi^2}{\Lambda^2 L^2}})}{L}
  \right]^3 \label{eq:theta3}
\end{equation}
where we have used $\vec{q}^{\,2}= q_x^2+q_y^2+q_z^2$ and that the exponential
of a sum is the product of the exponentials. This latter property allows to
relate the sum in three dimensions to the cube of the sum in one
dimension. In Eq.~\eqref{eq:theta3}, $\theta_3(u,\alpha)$ is a 
Jacobi elliptic theta function~\cite{theta3}. It satisfies~\cite{theta3bis},
\begin{equation}
\frac{\theta_3(0,e^{-\pi x^2})}{\theta_3(0,e^{-\pi/x^2})} = \frac{1}{x}~.
\end{equation}
This allows us to write then:
\begin{equation}\label{eq:derdG}
\frac{\partial\ \delta G}{\partial \Lambda} = - \frac{\mu}{(2\pi)^{3/2}} e^{2k^2/\Lambda^2} \left( \left[\theta_3(0,e^{-\Lambda^2 L^2/8})\right]^3 - 1 \right)~.
\end{equation}
We note that this equation is exact. The above equation converges
rapidly to zero as the Gaussian cutoff increases, which shows that
the limit $\Lambda \to \infty$ is effectively quickly achieved. To
proceed further, we note that:
\begin{equation}
\left[\theta_3(0,\alpha)\right]^3 = 1 + 6\alpha 
+ 12 \alpha^2 + \cdots = \sum_{m=0}^{\infty} c_{m} \alpha^m~,
\end{equation}
and the coefficients $c_m$ are nothing but the 
multiplicities of $m=\vec{n}^2$, $\vec{n} \in \mathbb{Z}^3$. 
Since $\alpha = e^{-\Lambda^2 L^2/8}$, we can find the 
leading term in Eq.~\eqref{eq:derdG}, 
\begin{equation}
\frac{\partial\ \delta G}{\partial \Lambda} = -
\frac{6\mu}{(2\pi)^{3/2}} \exp{\left( \frac{2k^2}{\Lambda^2} -
  \frac{\Lambda^2 L^2}{8}\right)}~\left(1 + {\cal O}\left (
e^{-\Lambda^2 L^2/8} \right) \right)~.
\end{equation}
Given that $\delta G = \delta G_L$ for $\Lambda\to\infty$, we find
keeping just the leading term:
\begin{align}
\delta G(E;\Lambda) & = \delta G_L(E) + \frac{6\mu}{(2\pi)^{3/2}}
\int_{\Lambda}^{\infty} \mathrm{d} \Lambda' \exp{\left(
  \frac{2k^2}{{\Lambda'}^2} - \frac{{\Lambda'}^2 L^2}{8}\right)} \\ 
& = \delta G_L(E) + \frac{3\mu}{2\pi L} \left[ e^{ikL}
  \mathrm{erfc}\left(\frac{\Lambda L}{2\sqrt{2}} + i
  \frac{\sqrt{2}k}{\Lambda} \right) + e^{-ikL}
  \mathrm{erfc}\left(\frac{\Lambda L}{2\sqrt{2}} - i
  \frac{\sqrt{2}k}{\Lambda} \right) \right]~, \nonumber
\end{align}
and, then, its asymptotic behavior is:

\begin{equation} \label{eq:asy_dG}
\delta G(E;\Lambda) = \delta G_L(E) + \frac{24\mu}{(2\pi)^{3/2}}
\frac{e^{-\frac{\Lambda^2 L^2}{8}}}{\Lambda L^2} \left[ 1 +
  \frac{2(k^2L^2-2)}{L^2\Lambda^2} + \mathcal{O}\left(
  \Lambda^{-4}\right) \right] + \cdots~,
\end{equation}
where $\mathcal{O}(\Lambda^{-4})$ refers to $(k/\Lambda)^4$, $(k^2/L^2)/\Lambda^4$ and
$1/(L\Lambda)^4$, and the ellipsis stands for terms that are more
exponentially suppressed (the next one would take the form
$e^{-\Lambda^2 L^2/4}$). Given the form of the $L$ suppression, the
Gaussian regularization scheme does  not introduce any spurious 
terms that would dominate
over\footnote{As already mentioned, for $k^2>0$ $\delta G_L$ 
is not exponentially suppressed for $L\to
\infty$, while for $k^2 < 0 $, we expect $\delta G_L$ to decrease as
$\exp(-|k|L)$.}  the  physical contribution $\delta G_L$, as long as
$\Lambda$ is sufficiently large. Indeed,  one can efficiently
compute the regularized L\"uscher function by means of the Gaussian
regulated $\delta G$ loop function. The L\"uscher 
function\footnote{It satisfies~\cite{Luscher:1990ux},
\begin{equation*}
e^{2i\delta} = \frac{k \cot\delta + ik}{k \cot\delta - ik} = 
\frac{{\cal Z}_{00}(1,\hat k^2)+ i \pi^\frac32 \hat k}{{\cal
    Z}_{00}(1,\hat k^2)- i \pi^\frac32 \hat k}.
\end{equation*}
}  
is related to the loop functions by means of~\cite{Doring:2011vk}: 
\begin{equation}
\sqrt{4\pi}{\cal Z}_{00}(1,\hat k^2) = -\frac{L}{2\pi}
  \frac{(2\pi)^3}{2\mu} \delta G_L(E), \quad \hat k^2 = \frac{k^2 L^2}{(2\pi)^2}~.
\end{equation}
Thus, for a mildly large value of
$\Lambda$, $\delta G_L(E)$ can be approximated by  the Gaussian
regulated $\delta G (E, \Lambda)$ function, up to corrections suppressed by the
exponential factor $e^{-\frac{\Lambda^2 L^2}{8}}$ (see Eq.~\eqref{eq:asy_dG}), 
\begin{equation}
\sqrt{4\pi}{\cal Z}_{00}(1,\hat k^2) = -\frac{L}{2\pi}
  \frac{(2\pi)^3}{2\mu} \left (\delta G(E, \Lambda)+\mathcal{O} \left(e^{-\frac{\Lambda^2 L^2}{8}}\right)\right)
\end{equation}
which in turn
provides ${\cal Z}_{00}(1,\hat k^2)$  with enough accuracy in a computationally easy way.

\section{Cutoff effects and relation to dispersion relations}
\label{app:parameters}

In this Appendix and to better frame the approach
followed in Subsec.~\ref{subsec:potential}, the existing 
correlation between the constant of the
potential and the cutoff is addressed in detail. We also discuss the
relation of our approach to other approaches in which the loop
function is calculated from a dispersion relation.

We recall Eqs.~\eqref{eq:gmat_gr} and \eqref{eq:potential_gr} to
expand the inverse of the amplitude in powers of $k^2$. For more
general purposes, we consider  a potential in which the factor that
multiplies the Gaussian, $\exp{(-2k^2/\Lambda^2)}$, has some energy
  dependence instead of being constant. It is to say, we replace in Eq.~\eqref{eq:potential_gr} $C$ by 
$C(E) = c_1 + c_2 k^2$, that reduces to the original form by setting
$c_2=0$. We find
\begin{equation}
V^{-1}-G = \frac{1}{c_1} + \frac{\mu\Lambda}{(2\pi)^{3/2}} +
 \left( - \frac{c_2}{c_1^2} + \frac{2}{c_1 \Lambda^2} -
 \frac{2\mu}{(2\pi)^{3/2}\Lambda} \right) k^2 + i\frac{\mu k}{2\pi} + \mathcal{O}(k^4)~.
\label{eq:correlac}
\end{equation}
For model-given values of $c_1$ and $c_2$ for an imposed cutoff
$\Lambda$, one can shift the cutoff to $\Lambda'$ and have the same
$T$-matrix, up to $\mathcal{O}(k^4)$, by reabsorbing the cutoff shift
in the new parameters $c'_1$ and $c'_2$, given by:
\begin{align}
\frac{1}{c'_1} & = \frac{1}{c_1} + 
\frac{\mu(\Lambda-\Lambda')}{(2\pi)^{3/2}}~,\label{eq:c1_GR}\\
\frac{c'_2}{c_1^{'2}} & = \frac{c_2}{c_1^2} + 
\frac{2}{c_1} \frac{\Lambda^2-{\Lambda'}^2}{\Lambda^2{\Lambda'}^2} 
+ \frac{2\mu(\Lambda-\Lambda')^2}{(2\pi)^{3/2}\Lambda {\Lambda'}^2}~.\label{eq:c2_GR}
\end{align}
If we insist in a constant potential, $c_2=c'_2=0$, we can also
reabsorb the cutoff effects in $c_1$, but this would be correct just
up to $\mathcal{O}(k^2)$.

Let us consider an approach in which the amplitude is written with a
loop function regularized by means of a once--subtracted dispersion
relation (DR),\footnote{The expression for the loop function can be
found by applying the dispersion relation integral, or, in a more
handy way, by taking the limit $\Lambda \to \infty$ for the case of
the Gaussian regulator loop integral and reabsorbing the infinity in
the subtraction constant.}
\begin{align}
T^{-1}_\text{DR} & = V^{-1}_\text{DR} - G_\text{DR}~,\\
V_\text{DR} & = C~,\\
G_\text{DR} & = \alpha - i\frac{\mu k}{2\pi}~,
\end{align}
where $C$ is the potential, analogous to the case of the Gaussian
regulator approach, and $\alpha$ is a subtraction constant (a free
parameter of the approach). Considering, as before, $C = c_1 + c_2
k^2$, we can expand:
\begin{equation}
V_\text{DR}^{-1} - G_\text{DR} = \frac{1}{c_1}-\alpha 
- \frac{c_2}{c_1^2}k^2  + i\frac{\mu k}{2\pi} + \mathcal{O}(k^4)~.
\end{equation}
We can then reabsorb the effects up to $\mathcal{O}(k^4)$ of an
arbitrary shift in the subtraction constant by means of:
\begin{align}
\frac{1}{c_1'} & = \frac{1}{c_1} + \alpha'-\alpha~,\\
c'_2 & = \frac{c_1^{'2}}{c_1^2}c_2~.
\end{align}
We see that the effects of the shift can be reabsorbed {\it exactly}
for a constant potential, with the first of the previous
equations. However, in the more general case of energy dependent
potentials (as the case of chiral potentials, for example), this
cannot be made exactly but just up to $\mathcal{O}(k^4)$. We see thus
that there are several equivalent ways: one can fit, in the Gaussian
regulator case, a constant for the potential and the cutoff, or fix
the latter to a reasonable (but otherwise arbitrary) value and fit two
constants. In a dispersion relation, a similar situation is found,
where now the subtraction constant plays the equivalent role of the
cutoff. In Subsec.~\ref{subsec:potential} we have
followed the first approach. In Fig.~\ref{fig:cutoff_correlation},
we show, for the case of $I=0$ $J^{PC}=0^{++}$ $D\bar{D}$, the
contours curves of the $\chi^2$ function in terms of the free
parameters: the UV cutoff $\Lambda$ and the inverse of the constant ($C_{0a}$)
that appears in  the potential for this channel.  We see already from this figure that
$\Lambda$ and $1/C_{0a}$ are strongly correlated, and that the
correlation is of the form given in Eq.~\eqref{eq:c1_GR}. Indeed, the
dashed line in the plot, that lies close to the axis of the error
ellipse, is Eq.~\eqref{eq:c1_GR} using  the central values of the cutoff
and the potential given in the original work of
Ref.~\cite{HidalgoDuque:2012pq}, $\Lambda=1000\ \text{MeV}$ and
$C_{0a} = -1.024\ \text{fm}^2$. We also infer that the correlation is
stronger for higher values of the cutoff, since the quadratic terms in
Eq.~\eqref{eq:correlac} become less important as $\Lambda$ increases.
\bibliographystyle{plain}

\end{document}